\documentclass[11pt,a4paper]{article}
\usepackage[utf8]{inputenc}
\usepackage{amsmath}
\usepackage{amsfonts}
\usepackage{amssymb}
\usepackage{amsbsy}
\usepackage{graphicx}
\usepackage{multirow}
\usepackage{caption}

\usepackage{threeparttable}
\usepackage{changepage}
\usepackage{placeins}

\usepackage{xcolor}
\usepackage{colortbl}

\usepackage{makecell}

\usepackage{tikz}
\usetikzlibrary{positioning,fit,arrows,shapes.geometric,chains,decorations.pathreplacing,arrows.meta}

\usepackage[super]{nth}

\usepackage{hyperref}
\hypersetup{pdfborder={0 0 1}}
\usepackage{adjustbox}

\usepackage[left=3cm,right=3cm,top=2.5cm,bottom=2.5cm]{geometry}
\usepackage{mathrsfs}

\usepackage{tikz}

\usepackage[american]{babel}
\usepackage{csquotes}

\usepackage[%
    style=numeric,
    natbib=true,
    giveninits=true,
    uniquename=init,
    alldates=year,
    backend=biber]{biblatex}
\DeclareLanguageMapping{american}{american-apa}

\newrobustcmd*{\parentexttrack}[1]{%
  \begingroup
  \blx@blxinit
  \blx@setsfcodes
  \blx@bibopenparen#1\blx@bibcloseparen
  \endgroup}

\title{\textbf{Two-Stage Sector Rotation Methodology Using Machine Learning and Deep Learning Techniques\thanks{We are very grateful to Satyan Malhotra, CEO at Ask2.ai, Inc as the Senior Advisor and Industry Expert. We would like to thank 
Susan	Chen, Annan	Chen, Modhuli	Goswami, Jitesh	Gurav, Sarin	Indrasukhsri, Nikhil	Kamoji, Michael	Pelton, Sarthak	Tiwari, Kin Wai	Wong, Ziyu	Zhang for their participation and help on this research. Errors are our own responsibility.
}}}
\date{}
\author{Tugce Karatas\footnote{Department of IEOR, Columbia University, \textbf{tk2757@columbia.edu}} \and  Ali Hirsa\footnote{Department of IEOR, Columbia University, \textbf{ah2347@columbia.edu}}}

\begin{document}

\maketitle
	
\begin{abstract}
Market indicators such as CPI and GDP have been widely used over decades to identify the stage of business cycles and also investment attractiveness of sectors given market conditions. In this paper, we propose a two-stage methodology that consists of predicting ETF prices for each sector using market indicators and ranking sectors based on their predicted rate of returns. We initially start with choosing sector specific macroeconomic indicators and implement Recursive Feature Elimination (\texttt{RFE}) algorithm to select the most important features for each sector. Using our prediction tool, we implement different Recurrent Neural Networks (\texttt{RNN}) models to predict the future ETF prices for each sector. We then rank the sectors based on their predicted rate of returns. We select the best performing model by evaluating the annualized return, annualized Sharpe ratio, and Calmar ratio of the portfolios that includes the top four ranked sectors chosen by the model. We also test the robustness of the model performance with respect to lookback windows and look ahead windows. Our empirical results show that our methodology beats the equally weighted portfolio performance even in the long run. We also find that Echo State Networks (\texttt{ESN}) exhibits an outstanding performance compared to other models yet it is faster to implement compared to other \texttt{RNN} models. 

\end{abstract}

\providecommand{\keywords}[1]{\textbf{\textit{Keywords:}} #1}

\section{Introduction}
 
US economy has been following business cycles throughout the history. There are 11 business cycles identified by Natural Bureau of Economic Research since 1945 \footnote{The details of these business cycles are documented in \url{https://www.nber.org/research/data/us-business-cycle-expansions-and-contractions}}. Each business cycle has its own characteristics, yet there are similar patterns observed in each cycle. A typical business cycle starts with an economic growth until reaching a peak, and continues with an economic recession till attaining a trough. Market indicators have been used as the key determinants at identifying the stages of business cycles over decades. As a well-known model released by Merrill Lynch in 2004, Investment Clock uses growth (GDP) and inflation (CPI) factors to identify the four stages of the business cycle. 


It has been observed that sector rotations occur at each stage of the business cycle. Depending on the phase of the cycle, investments on certain sector stocks tend to be more appealing. Sector rotation strategies have been built as a top-down approach to take advantage of the changing attractiveness of the sectors through business cycles. The stage of the business cycle is identified by analyzing macroeconomic indicators, and then investments are made based on the outperforming sectors for the chosen cycle. Numerous studies have shown that sector rotation strategies mostly beat the performance of the overall market.

Investment Clock is a very intuitive model, yet it only uses two factors to identify the status of the economy. In this paper, we benefit from a large number of common and sector specific macroeconomic indicators to identify the outperforming sectors as time progress. We aim at building a sector ranking model based on relative growth potential and investment attractiveness of eight major sectors: healthcare, technology, industrials, utilities, materials, energy, financial services and consumer discretionary. The ranking model will then be used for guiding private equity funds in the future in terms of sector targeting at particular points in time. We use public market information (ETFs) as a proxy for private market data because there is no available sector-wise private equity data, and the publicly available data for private equity funds is limited and scarce. 

The contribution of this paper to the literature is two-fold. First, the existing literature is very limited on sector ranking models that are built using macroeconomic variables. In our paper, we consider a variety of sector-specific macroeconomic indicators for each sector and only add the most meaningful features into each sector prediction model. To the best of our knowledge, current literature only focuses on predicting sector rankings one month ahead. As our second contribution, we predict the sector rankings over different time horizons, and validate that our methodology provides comparable results even in the long run.

This paper is organized as follows: In Section \ref{LR-MI}, we go through the existing literature on market indicators. In Section \ref{Data-MI}, we describe our data set and the data preprocessing techniques that we implement for preparing data for modeling. Section \ref{Preliminary-MI} provides a technical background on the models that we implemented in this paper. We introduce our overall sector ranking methodology in Section \ref{Methodology-MI}. Section \ref{Results-MI} starts with introducing the performance measures that are used in this paper. We also provide the backtesting results and the evolution of performance measures with increasing lookback window.  Finally, we summarize our initial findings and explain our future extension plans in Section \ref{Conclusion - MI}.

\section{Literature Review} \label{LR-MI}
This paper contributes to two strands of literature: use of macroeconomic variables in asset return forecasting and building sector rotation methodologies. In Section \ref{macro}, we introduce papers that use macroeconomic variables as predictors in asset return forecasting. Section \ref{sector_rotation} overviews the literature on tactical asset allocation using sector rotation strategies.

\subsection{Macroeconomic Variables in Asset Return Forecasting} \label{macro}
Numerous studies have attempted to identify the macroeconomic variables to be associated with the asset return prediction in different markets. Chen et al. \cite{chen1986economic} experimented with seven macroeconomic variables to explain the stock returns in US stock markets, and showed that industrial production, changes in risk premia, twist in yield curve, and inflation affect stock prices. On the other hand, they found that consumption and oil price do not have explanatory power on stock price behavior. In another study, Chen \cite{chen1991financial} found that default spread, term spread, one month T-bill rate, industrial production growth rate and dividend price ratio significantly impact stock prices. Clare and Thomson \cite{clare1994macroeconomic} analyzed the impact of 18 macroeconomic variables in UK stock markets, and concluded that oil price, retail price index, bank lending, and corporate default risk have impact on stock price movement. Mukherjee and Naka \cite{mukherjee1995dynamic} identified the relationship between Japanese stock returns and exchange rate, inflation rate, money supply, real economic activity, long-term government bond rate, and call money rate. With a similar study in Norway stock markets, Gjerde and Saettem \cite{gjerde1999causal} ascertained the positive correlation of stock returns with oil price and real economic activity. They failed to show a strong correlation between stock returns and inflation. Chung and Shin \cite{kwon1999cointegration} investigated the effect of macroeconomic variables on Korean stock prices, and revealed the significant impact of trade balance, foreign exchange rate, industrial production, and money supply on the stock prices. In another US market study, Flannery and Protopapadakis \cite{flannery2002macroeconomic} investigated the impact of balance of trade, housing starts, employment, CPI, M1, and producer price index on equity trading volume. Ibrahim and Aziz \cite{ibrahim2003macroeconomic} found that CPI and industrial production are positively related to Malaysian stock prices in the long-run. They also observed that stock prices have negative associations with exchange rate and money supply. Adam et al. \cite{adam2008macroeconomic} analyzed the short-term and the long-term effects of macroeconomic indicators on Ghanaian stock market indexes using Johansen's multivariate cointegration test and innovation accounting techniques. Their experiments reveal that stock market indexes are significantly affected from inflation and exchange rates in the short-run, and from interest rate and inflation in the long-run. Singh et al. \cite{singh2011macroeconomic} found that exchange rate and GDP affects overall index returns in Taiwanese stock market. They showed that inflation rate, exchange rate, and money supply have negative associations with portfolios constructed using stocks of medium and large companies. Chong et al. \cite{chong2012eta} introduced a macroeconomic factor model, Eta model, using 18 macroeconomic variables to predict stock returns. In their model, they only included systematic variables such as unemployment rate, CPI, energy prices, and M2 money supply. They compared the performance of their model with that of Fama-French three-factor model. The experimental results show that their macroeconomic factor model outperforms the famous fundamental factor model. Chong and Phillips \cite{chong2014tactical} implemented the Eta model on tactical asset allocation. They defined a metric called ECR (economic climare rating) that scores the impact of current economy on ETFs. The ratings change from one to five with five means economic environment is favorable for that ETF. Their strategy of using ETFs with 3, 4, and 5 ratings, and applying Mean-Variance Optimization outperformed benchmark models. Jare$\tilde{n}$o and Negrut \cite{jareno2016us} analyzed the impact of GDP, CPI, IPI, unemployment rate, and long-term interest rate on US stock prices, and found that all macroeconomic variables have significant effects except CPI. They further showed the positive association of GDP and IPI, and the negative association of unemployment rate and interest rate with stock prices. Misra \cite{misra2018investigation} investigated the movement of Indian stock market using market indicators, and found a long-term impact of IPI, inflation, interest rate, gold price, exchange rate, foreign institutional investment, and money supply, and a short-term impact of inflation and money supply based on Vector Error Correction Model.

\begin{table}[!htbp]
\vskip\baselineskip 
\begin{center}
\begin{adjustwidth}{-0.8cm}{0cm}
\scalebox{0.65}{
\begin{tabular}{c c || c || c|c|c} \hline
\textbf{Paper} & \textbf{Year} & \textbf{Stock Market} & \textbf{Significant Market Indicators} & \textbf{Positive Association} & \textbf{Negative Association} \\ \hline
\cite{chen1986economic} & 1986 & US & \makecell{industrial production, twist in yield curve, \\changes in risk premia, inflation} & - & -  \\\hline
\cite{chen1991financial} & 1991 & US& \makecell{default spread,one month T-bill rate, \\ term spread, dividend price ratio\\ industrial production growth rate} & - & - \\ \hline
\cite{clare1994macroeconomic} & 1994 & UK & \makecell{oil price, retail price index, bank lending \\ corporate default risk} & - & - \\\hline
\cite{mukherjee1995dynamic} & 1995 & Japanese & \makecell{inflation,real economic activity, \\money supply, call money rate\\ long-term government bond rate} & -&-\\\hline
\cite{gjerde1999causal} & 1999 & Norway & oil price, real economic activity & oil price, real economic activity & - \\ \hline
\cite{kwon1999cointegration} & 1999 & Korean &\makecell{trade balance, foreign exchange rate \\ industrial production, money supply} &-&-\\ \hline
\cite{flannery2002macroeconomic} & 2002 & US & \makecell{trade balance, housing starts, \\ employment, CPI, M1, PPI} & - & - \\ \hline
\cite{ibrahim2003macroeconomic}&2003&Malaysia & \makecell{industrial production, CPI, money supply, \\ exchange rate} & industrial production, CPI & money supply, exchange rate \\ \hline
\cite{adam2008macroeconomic} & 2008 & Ghana & \makecell{inflation, exchange rate, interest rate} & - & - \\ \hline 
\cite{singh2011macroeconomic} & 2011 & Taiwan & \makecell{exchange rate, GDP, inflation rate, \\money supply} & - & \makecell{inflation rate, exchange rate,\\ money supply} \\ \hline 
\cite{jareno2016us} & 2016 & US & \makecell{GDP, IPI, unemployment rate, \\ long-term interest rate} & GDP, IPI & unemployment rate, interest rate \\ \hline 
\cite{misra2018investigation} & 2018 & Indian & \makecell{IPI, inflation, interest rate, \\ gold price, exchange rate, money supply, \\ foreign institutional investment} & - & -\\ \hline
\end{tabular} }
\end{adjustwidth}
\end{center}
\caption{Literature on Macroeconomic Variables Used in Stock Return Forecasting}
\label{table:literature}
\end{table}
\FloatBarrier

Table \ref{table:literature} summarizes the literature associated with the impact of macroeconomic variables on different markets. Although they are built using different market data, some market indicators have significant impacts on multiple markets such as inflation rate, CPI, IPI, and GDP. The observations indicate that unemployment rate, interest rate, and exchange rate are negatively impacting stock market movements, which is intuitively reasonable. Similarly, GDP, CPI, IPI, and oil price are positively associated with stock prices. The literature on macroeconomic variables solely focuses on the overall market stock prices. To the best of our knowledge, there is no existing literature that analyzes the sector-level impact of macroeconomic variables. In our paper, we first identify the possible sector-specific macroeconomic variables, and then investigate their impact on sector index returns based on US stock market.

\subsection{Sector Rotation} \label{sector_rotation}
A number of studies in the literature have constructed sector rotation strategies to outperform the performance of the overall market. There are two different research areas for sector rotation strategies. In the first research area, sector indexes are predicted using a variety of variables, and then sectors are ranked based on their predicted returns. In the second research field, the phase of the business cycles are detected, and investment strategies are developed based on outperforming stocks within the predicted phase.

The first research area focuses on building prediction models for sector indexes, and then rank sectors based on their returns. Moskowithz and Grinblatt \cite{moskowitz1999industries} observed that investment momentum strategy of buying stocks from winning industries and selling stocks from losing industries is profitable. In their model, they rank 20 US sector stocks based on their six month historical stock prices, and then invest in top 30 percent of stocks while shorting low 30 percent of stocks. Chong and Phillips \cite{chong2015sector} implemented the macroeconomic Eta model on sector rotation strategies. The model is calibrated using three-years data with 18 macroeconomic factors to predict ETF sector indexes. They selected sectors based on the criteria of 95 percent or higher $R^2$ value in Eta model, and applied Mean Variance Optimization (\texttt{MVO}) to construct the portfolio. By following long-only strategy and re-balancing the portfolio semiannually, they found that their strategy outperforms benchmark portfolios. Gao and Ren \cite{gao2015new} employed principal component regression to predict future sector indexes in Shanghai Security Market. They calibrated their model using four years data, and predicted the sector returns for the following week. Their strategy of buying the top sector index and selling the lowest sector index achieved better average weekly returns in comparison to market composite index. Zhu et al. \cite{zhuutilizing} applied explainable \texttt{AI} models such as linear regression, ridge regression and random forest using five macroeconomic indicators (growth factor, inflation factor, rate factor, credit factors, and exchange factor), and predicted the monthly sector index returns in Chinese stock market. The models are calibrated using two years data, and the next month returns are predicted. They experimented with one month, six months, and one year lagged indicators, and implemented \texttt{PCA} and feature importance with random forest on lagged indicators. They selected top five sectors based on their rate of return, and their monthly re-balanced portfolio outperformed equally-weighted index portfolio.

There also exists studies on prediction models for the phase of the business cycle. The investment strategies are built based on the performance of sectors within predicted stage of the business cycle. Greetham and Hartnett \cite{greetham2004investment} introduced Investment Clock in 2004 in order to identify four stages of a business cycle: recovery, overheat, stagflation, and reflation. Different stages of the business cycle are detected based on the trend of growth (GDP) and inflation (CPI). They observed that consumer discretionary, telecom, and technology stocks perform well during recovery; whereas technology, industrials, oil and gas sector stocks perform the best during overheat. Outperforming stocks in stagflation phase are the stocks from oil and gas, pharmaceuticals, and utilities sectors. Consumer staples, financials, and consumer discretionary stock prices are tend to be high over reflation period. Raffinot and Benoit \cite{raffinot2018investing} implemented random forest and boosting algorithms to predict the turning points in the economy based on the deviation of real GDP from its trend. Their model consists of two types of turning points: acceleration and slowdown. According to their investment strategy, they invest 80\% of their money to equities, and the remaining 20\% to the bonds when acceleration is expected in the economy. Otherwise, 40\% of the portfolio will belong to equities, and the rest will be invested in bonds. Sauer \cite{sauer2019sector} divided the status of economy into four cases based on the regimes observed in GDP: high rising, high falling, low rising, and low falling. Since GDP is reported quarterly, the data is transformed into monthly frequency using monthly industrial production as an indicator. Two independent random forest models are developed to predict two-level regime (high vs low) and two-level momentum (rising vs. falling). By excluding cyclical sectors in low falling regime, they managed to outperform equally-weighted benchmark. Wang et al  \cite{wang2020novel} divided economy into four regimes based on GDP year over year change and 10 year treasury bond yield. They implement k-Nearest Neighbors algorithm to predict the status of the economy based on macroeconomic factors. Furthermore, they employed post-Lasso regression for all sector indexes in each economic regime to predict the next month sector returns using the related sector historical return data. Based on the status of the economy, corresponding post-Lasso models are used to predict sector returns, and the strategy of longing top 20\% sector indexes and shorting the bottom 20\% of sector indexes is employed. Their investment strategy mostly outperformed the equally-weighted benchmark. 

In our paper, we implement \texttt{ML} \& \texttt{DL} models to predict sector index returns using sector-specific macroeconomic indicators. We then rank the sectors based on their predicted rate of returns, and select top four ranked sectors for long-only strategy. Our paper belongs to the first research area, yet the existing literature focuses only on the one month ahead predictions. We extend our prediction horizon up to two years, and evaluate the robustness of our methodology by experimenting with different prediction horizons. 

\section{Data} \label{Data-MI}
To compare the performance of different sectors, we utilize iShares ETF prices for eight major sectors: healthcare (\texttt{IYH}), energy (\texttt{IYE}), utilities (\texttt{IDU}), finance (\texttt{IYG}), technology (\texttt{IYW}), materials (\texttt{IYM}), industrials (\texttt{IYJ}), and consumer goods (\texttt{IYK}). Our dataset consists of ETF prices for each sector from July 14, 2000 to Nov 10, 2019. As we use the market calendar of NYSE, we obtain 4,862 daily adjusted close prices for each ETF. We utilize yfinance python library to download the data. In this paper, we mostly experiment with monthly data. Therefore, we obtain 233 monthly adjusted close prices. 

Macroeconomic variables that we consider in this study comes from a variety of resources. Appendix \ref{macroeconomic} provides the tables of macroeconomic variables for each sector together with their resources and data frequencies. Table \ref{table:all_sectors} gives the list of macroeconomic variables that are commonly used as input for each sector model. These macroeconomic variables include gross domestic product (\texttt{GDP}), unemployment rate, consumer price index (\texttt{CPI)}, 30-year fixed rate mortgage rate, and federal funds rate. As it is seen, macroeconomic variables are reported on different time frequency basis varying from daily to annually. Thus, we implement linear interpolation to obtain monthly observations for each macroeconomic variable. After the interpolation, the dataset of macroeconomic variables align with sector ETF prices. 

For each sector, we define the relevant macroeconomic variables, but using too many features leads to an increase in the model complexity, and it may further cause multicollinearity between variables and model overfitting. Recursive Feature Elimination (\texttt{RFE}) (Guyon et al. \cite{guyon2002gene}) algorithm is implemented to avoid these problems and to obtain the most important features for each sector. \texttt{RFE} is a backward selection algorithm, which recursively fits a machine learning model using all available features, and calculates the importance of each feature within the model. The algorithm eliminates the least important features at each iteration until the predetermined number of features are obtained. The relative importance of each feature is measured by the average decrease in node impurity caused by that feature. In this paper, we employ random forest regression as a sub-routine machine learning model, and we select top four ranked features as the final set of features.

\begin{figure}[!htbp]
\captionsetup{font=scriptsize,labelfont=scriptsize}
   \begin{minipage}{0.5\textwidth}
   \hspace*{-2.0cm}
     \centering
     \includegraphics[scale = 0.35]{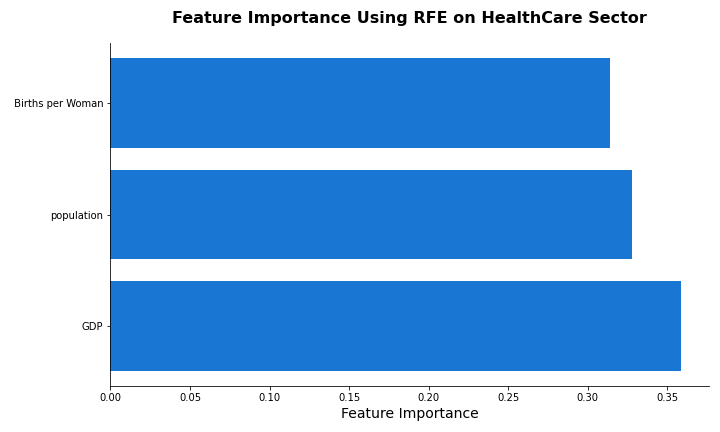}
   \end{minipage}\hfill
   \begin{minipage}{0.5\textwidth}
   \hspace*{0.1cm}
     \centering
     \includegraphics[scale = 0.35]{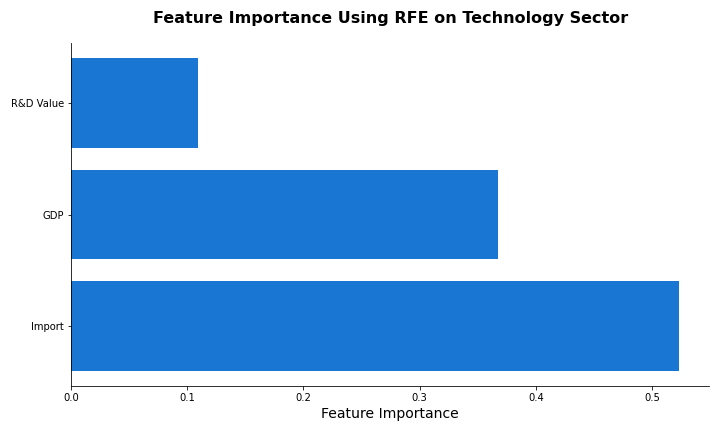}
   \end{minipage}
 \caption{Feature Importance Using RFE on Different Sectors} \label{fig:rfe}
\end{figure}
\FloatBarrier

Figure \ref{fig:rfe} provides feature importance plots obtained using \texttt{RFE} algorithm for finance and technology sectors. According to the plots, GDP is a very significant feature while predicting sector index prices. Finance sector ETF prices are highly affected by unemployment rate and trade balance \% of GDP. R\&D value and import are also key drivers for technology sector index prices which is expected by intuition.

\section{Preliminaries} \label{Preliminary-MI}

In this section, we provide a background on the existing machine learning and deep learning models that we implement within our overall model framework. These models are ridge regression, long short-term memory (\texttt{LSTM}), gated recurrent unit (\texttt{GRU}), and echo state networks (\texttt{ESN}), respectively.

\subsection{Ridge Regression} \label{rr}
Ridge regression is a type of regularized regression proposed by Hoerl and Kennerd \cite{hoerl1970ridge2},\cite{hoerl1970ridge} to control the high variance of estimates due to the multicollinearity between the features. We apply ridge regression as our benchmark model because it is commonly used in industry. Ridge regression estimates $\beta$ regression coefficients within the formula:
\begin{equation*}
    y=X\beta+\epsilon,
\end{equation*}
where $X\in\mathbb{R}^{n\times p}$ is the predictor matrix, $y\in\mathbb{R}^n$ is the response vector, and $\epsilon \in \mathbb{R}^n$ is residuals. Although very similar to linear regression, ridge regression adds $L_2$ penalty term on the coefficients in the objective function, and controls the strength of penalty with $\lambda$ tuning parameter. In the existence of multicollinearity, the variance of the estimates will be high. By adding penalty, ridge regression shrinks the regression coefficients towards zero and controls the variance of estimates. Equation \ref{ridge} provides the analytical solution for $\beta$ coefficients. 
\begin{equation}
    \beta_{ridge} = \arg \min_{\beta} ||y-X\beta||_2^{2}+\lambda||\beta||_2^{2}
    =(X^\intercal X+\lambda I)^{-1}X^\intercal y \label{ridge}
\end{equation}
In comparison to linear regression where there is an uncertainty on non-singularity of $X^\intercal X$, $X^\intercal X+\lambda I$ in ridge regression is a non-singular and hence an invertible matrix that stabilizes the estimates. Despite the certain advantages of ridge regression over linear regression, we expect that the performance of \texttt{RNN}s will be superior to that of ridge regression because the latter one do not consider the temporal patterns within the data.

\subsection{Recurrent Neural Networks (\texttt{RNN}s)} \label{rnn}
Vanilla feedforward neural networks (\texttt{FNN}s) do not consider any temporal dependencies within the data because they rely on the assumption of independence between inputs and outputs. They are not able to understand the dependencies within the time-series data. On the other hand, recurrent neural networks (\texttt{RNN}s) (Rumelhart et al. \cite{rumelhart1986learning}) carry a memory through time steps of the data to remember the temporal dynamic behavior of the data. \texttt{RNN}s can learn temporal dependencies within the data using their hidden units $h_t$. Equation \ref{rnn_eq} shows the general structure of the relationship between hidden units, where $f$ represents a non-linear activation function. At each time step, the values of hidden units are calculated based on a non-linear function $f$, and $f$ takes the values of hidden units at previous time steps as input together with the input $x_t$ from the current time step. 
\begin{equation}
    h_t = f(h_{t-1},x_t) \label{rnn_eq}
\end{equation}
\texttt{RNN}s are trained using Backpropagation Through Time (\texttt{BPTT}) algorithm. Hidden units in the same layer share the same weight matrices and bias vectors. \texttt{BPTT} algorithm calculates the gradient of chosen loss function with respect to these weight matrices and bias vectors. When the sequence of the data is long, vanishing gradient problem may occur. In the case of vanishing gradients, weight matrices and bias vectors are not updated enough and learning stops. Although \texttt{RNN}s are good at capturing short-term dependencies, they are vulnerable to failure at memorizing longer term relationships. Gated recurrent neural networks has been built to address the vanishing gradients problem and to understand long-term dependencies within the data. In this study, we benefit from two most popular gated \texttt{RNN}s: Long Short-Term Memory (\texttt{LSTM}) and Gated Recurrent Units (\texttt{GRU}s).  

\subsection{Long Short-Term Memory (\texttt{LSTM})} \label{lstm}
\begin{figure}[!htbp]
\centering
\hspace*{0.5cm}
\includegraphics[scale=0.4]{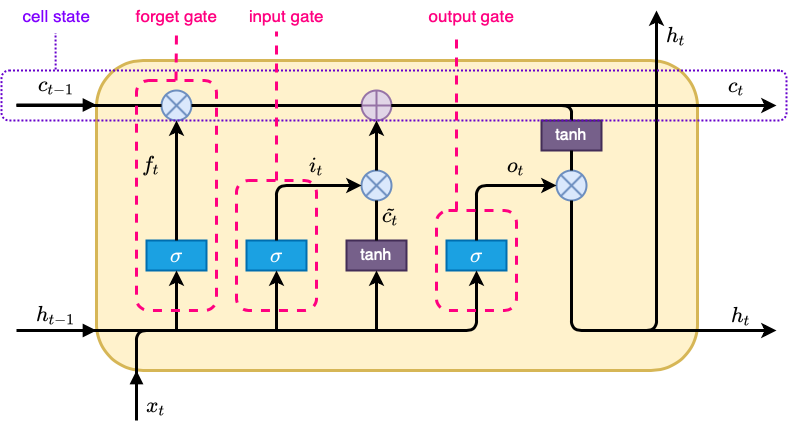}
\captionof{figure}{LSTM Memory Cell}
\label{plot:lstm}
\end{figure}

Long Short-Term Memory (\texttt{LSTM}) (Hochreiter and Schmidhuber \cite{hochreiter1997long}) is a type of \texttt{RNN}s, where $f$ non-linear activation function in Equation \ref{rnn_eq} is replaced by an \texttt{LSTM} memory cell. Each memory cell controls the flow of memory cell state $c_t$ by adding three gates: forget gate $f_t$, input gate $i_t$, and output gate $o_t$. Figure \ref{plot:lstm}
illustrates a typical \texttt{LSTM} memory cell. 

\FloatBarrier

As it is seen in Figure \ref{plot:lstm}, each input gate takes hidden state from previous time step $h_{t-1}$ and new information from current time step $x_t$ as input. Sigmoid activation function $f(x)=\frac{1}{1+e^{-x}}$ is then used to control the level of information flow in each gate (Equations \ref{f_lstm}, \ref{i_lstm}, and \ref{o_lstm}). While forget gate $f_t$ determines the amount of existing memory $c_{t-1}$ to forget, input gate $i_t$ decides on how much to add a new memory from $\Tilde{c_t}$ into the current cell state (Equation \ref{c_lstm}). Output gate $o_t$ calculates which parts of cell state $c_t$ to keep in hidden state $h_t$.

\begin{align}
    f_t &= \sigma\left(W_f \cdot h_{t-1} + U_f \cdot x_t + b_f\right) \label{f_lstm}\\
    i_t &= \sigma\left(W_i \cdot h_{t-1} + U_i \cdot x_t + b_i\right) \label{i_lstm}\\
    \Tilde{c_{t}} &= tanh\left(W_c \cdot h_{t-1} + U_c \cdot x_t + b_c\right) \label{tildec_lstm}\\
    c_t &= f_t \circ c_{t-1} + i_t\circ \Tilde{c_{t}} \label{c_lstm} \\
    o_t &= \sigma\left(W_o \cdot h_{t-1} + U_o \cdot x_t + b_o\right) \label{o_lstm}\\
    h_t &= o_t \circ tanh(c_t)\label{h_lstm}
\end{align}

\texttt{LSTM} memory cell flow is regularized by sigmoid ($\sigma$) and hyperbolic tangent ($tanh$) activation functions. Since $\sigma$ takes values between 0 and 1, and $tanh$ takes values between -1 and 1, the flow of information is controlled by bounded non-linear functions. Therefore, \texttt{BPTT} algorithm works smoothly without observing vanishing gradients problem quickly. 

\subsection{Gated Recurrent Unit (\texttt{GRU})} \label{gru}

\begin{figure}[!htbp]
\centering
\hspace*{0.5cm}
\includegraphics[scale=0.4]{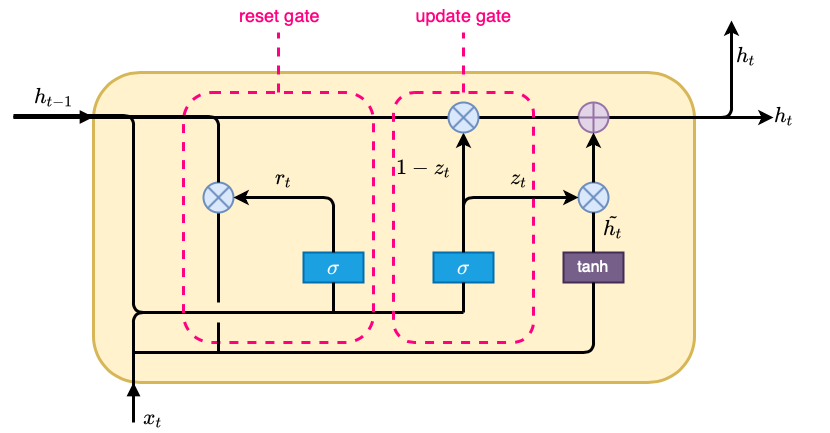}
\captionof{figure}{GRU Memory Cell}
\label{plot:gru}
\end{figure}

\texttt{GRU} is a more recent version of gated \texttt{RNN}s, which is introduced by Cho et al. \cite{cho2014learning} in 2014. Similar to \texttt{LSTM}, $f$ function in Equation \ref{rnn_eq} is replaced by a \texttt{GRU} unit. Each \texttt{GRU} unit consists of two gates: update gate $z_t$, and reset gate $r_t$. As opposed to \texttt{LSTM} memory cells, \texttt{GRU} units do not have a separate cell state. All information flow through hidden states $h_t$. In addition, forget gate and input gate in \texttt{LSTM} memory cells are represented with a single update gate $z_t$ in \texttt{GRU} units. Figure \ref{plot:gru} illustrates a typical \texttt{GRU} unit. 

\FloatBarrier
Equations \ref{z_gru} and \ref{r_gru} shows that each gate takes the previous hidden state $h_{t-1}$ and the new information from the current time step $x_t$ as input, and apply a sigmoid activation function to control the information flow. Reset gate decides on the extent to remember the previous information $h_t$. Update gate $z_t$ determines the level of information to forget from the previous hidden state $h_{t-1}$ and the level of information to add from the new candidate hidden state $\Tilde{h_t}$ (Equation \ref{h_gru}). 
\begin{align}
    z_t &= \sigma\left(W_z \cdot h_{t-1} + U_z \cdot x_t + b_z\right) \label{z_gru}\\
    r_t &= \sigma\left(W_r \cdot h_{t-1} + U_r \cdot x_t + b_r\right) \label{r_gru}\\
    \Tilde{h_{t}} &= tanh\left(W_h \cdot (r_t\circ h_{t-1}) + U_h \cdot x_t + b_h\right) \label{tilde_h_gru}\\
    h_t &= (1-z_t) \circ h_{t-1} + z_t\circ \Tilde{h_{t}} \label{h_gru}
\end{align}
\texttt{GRU} unit is also regularized by sigmoid and $tanh$ activation functions. Therefore, \texttt{BPTT} algorithm works smoothly and avoids vanishing gradients problem to a large extent. In their empirical study, Chung et al. \cite{chung2014empirical} compared the performance of \texttt{LSTM} and \texttt{GRU}, and they observed that both models are comparable to each other. \texttt{GRU} is faster than \texttt{LSTM} because it has fewer parameters in terms of number of gates and inexistence of the cell state. However, fewer number of parameters in \texttt{GRU} may also result in a decrease in the expressability. Since our data is limited, we expect that \texttt{GRU} model will be more suitable for our study compared to \texttt{LSTM} model.

\subsection{Echo State Networks (\texttt{ESN})} \label{esn}
\begin{figure}[!htbp]
\centering
\hspace*{0.5cm}
\includegraphics[scale=0.5]{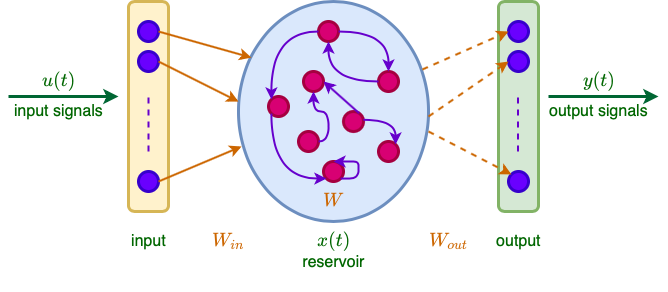}
\captionof{figure}{Echo State Networks}
\label{plot:esn}
\end{figure}

Echo State Networks (\texttt{ESN}s) (Jaeger \cite{jaeger2001echo}) are also a type of \texttt{RNN}s, but they address the vanishing gradients problem by employing a reservoir computing framework instead of gates. The model consists of three weight matrices: weights from input layer to reservoir ($W_{in}$), weights within reservoir ($W$), and weights from reservoir to output layer ($W_{out}$). $W_{in}$ and $W$ are randomly initialized, but they remain fixed throughout the training. $W_{out}$ is the only weight matrix trained to capture the temporal dynamics of the data. Figure \ref{plot:esn} illustrates the general architecture of \texttt{ESN}s.

\FloatBarrier

The algorithm consists of two steps. In the first step, the input signals $u(t)$ are converted into high-dimensional non-linear embeddings $x(t)$ through reservoir as in Equation \ref{esn_update_1}. The non-linearity of $x(t)$ comes from the function $f$, that is a non-linear activation function such as sigmoid and $tanh$. The relationship between $x(t)$ and output signals $y(t)$ are then calculated based on Equation \ref{esn_update_2}. Function $g$ is usually taken as linear function. Hence, a linear regression algorithm is trained to find $W_{out}$. Equations \ref{esn_update_1} and \ref{esn_update_2} are taken from Jaeger et al. \cite{jaeger2007optimization} because leaky integrators and tunable hyperparameters are included in the formulation. 

\begin{align}
    x(t+1) &= (1-\alpha)x(t)+f(s_{in}W_{in}u(t+1)+(\rho W)x(t)) \label{esn_update_1} \\
    y(t+1) &= g(W_{out}[x(t);u(t)]) \label{esn_update_2}
\end{align}

In Equation \ref{esn_update_1}, $\alpha$, $s_{in}$, and $\rho$ represent leaking rate, input scaling, and spectral radius, respectively. Leaking rate $\alpha$ determines the speed of the changes in $x(t)$. Input scaling $s_{in}$ is used to scale the weight matrix $W_{in}$. Spectral radius $\rho$ is the maximum absolute eigenvalue of the weight matrix $W$. The weight matrix $W$ is scaled by $\rho$, and larger spectral radius is used to consider longer memory. These hyperparameters together with reservoir size $N$ should be tuned in order to obtain the best model given the data.

Despite traditional \texttt{RNN}s, \texttt{ESN}s do not suffer from vanishing gradients problems as \texttt{ESN}s are not trained through \texttt{BPTT}. In addition, training of \texttt{ESN}s are faster compared to other \texttt{RNN}s because only a regression-like algorithm is used to train the model. Considering these two advantages, we expect that the performance of \texttt{ESN}s will be superior compared to the performance of other models we implement in this study.

\section{Proposed Methodology} \label{Methodology-MI}
\begin{figure}[!htbp]
    \centering
    \hspace*{-1.6cm}
    \includegraphics[scale=0.65]{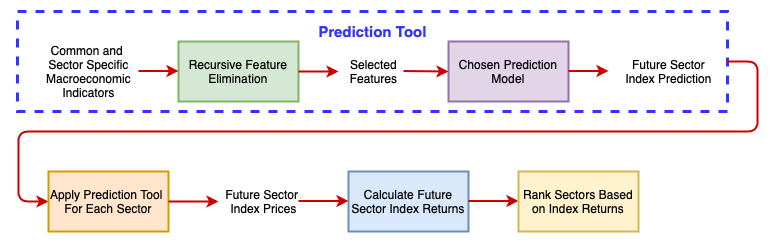}
    \caption{Overall Ranking Methodology}
    \label{fig:overall_ranking}
\end{figure}

In this section, we introduce our overall methodology for ranking sectors based on the market environment. Figure \ref{fig:overall_ranking} illustrates our overall methodology diagram. As in Figure \ref{fig:overall_ranking}, our methodology consists of two stages: prediction tool and ranking sector based on index returns. The prediction tool is applied on each sector independently. For each sector, we take common and sector specific macroeconomic indicators as input, and index prices as output. We employ Recursive Feature Elimination (\texttt{RFE}) on these variables, and pick the most important variables for each sector. Using selected features, we then predict the future sector index prices by employing the models from Section \ref{Preliminary-MI}. We implement our prediction tool for each sector, and obtain the predictions for the future sector index prices. In the second stage, we calculate future sector index returns using predicted sector index prices. Finally, we rank the sectors based on their predicted rate of returns. 

\FloatBarrier

\FloatBarrier

\section{Experimental Results} \label{Results-MI}

To evaluate the ability of our methodology on sector ranking that is introduced in Section \ref{Methodology-MI}, we conduct experiments to predict near-term, mid-term, and long-term rankings of sectors. In this section, we show our results for each prediction time horizon and investigate the behavior of different models with the change of lookback window. We also build a strategy for choosing the best performing model and lookback window.

We use the same values for the hyperparameters through all prediction models to analyze the robustness of the models across different look ahead periods. These values are obtained based on our earlier experiments. We use $\alpha=10$ for all ridge regression models. \texttt{LSTM} model is built using three hidden layers. The layers consist of 16, 256, and 64 hidden units, respectively. Our experiments show that \texttt{relu} activation function performs the best for all hidden layers. During the training, we try to minimize Root of Mean Squared Error (\texttt{RMSE}) using Adam optimizer with learning rate$=0.0001$ and decay$=1e-7$. We train the models for 1,000 epochs, and monitor the value of \texttt{RMSE} during the training for early stopping. The layout of \texttt{GRU} is similar to that of \texttt{LSTM}. We again use three hidden layers with \texttt{relu} activation function. Each hidden layer consists of 32, 256, and 64 hidden units, respectively. We train our models for 500 epochs, and allow for early stopping. \texttt{ESN} model is built using 100 reservoir units. The leaking rate, spectral radius, and reservoir density are set to 0.5, 1, and 0.5, respectively. For the regression routine of \texttt{ESN} model, we use ridge regression with $\alpha=1$. Finally, the transient time is chosen to be zero. It is important to note that these hyperparameters are chosen based on our preliminary experiments. We plan to integrate hyperparameter optimization techniques into our methodology as a future work. 

After running all models with a variety of lookback windows for each prediction horizon, we calculate the returns of each sector. In order to backtest the performance of different models on identifying sector rankings, we select top four sectors based on returns at each time step, and form an equally weighted portfolio based on the returns predicted through each model. We then compare the performance of the models with respect to each other and the benchmark portfolio. The benchmark portfolio is formed by all sector ETFs, and each sector shares an equal weight within the portfolio. As performance measure, we use annualized returns, annualized Sharpe ratio, and annualized Calmar ratio. 

\subsection{Performance Metrics}
As stated earlier, we use annualized return, annualized Sharpe ratio, and annualized Calmar ratio to evaluate the performance of portfolios that are created by using different models and benchmark portfolio. In this section, we briefly explain what they are. 

Annualized rate of return is the annual return of a portfolio over a given period of time. Using annualized rate of return, an investor can compare the performance of portfolios that are invested for different duration. Equation \ref{annualized_return} shows that annualized return is calculated as the geometric average of the returns over a given time horizon. Return in Equation \ref{annualized_return} is the percentage change of the price over given period. n indicates the period frequency within a year, and N stands for the total number of periods within given time horizon. In this paper, we mostly use n$=12$ because we consider monthly returns. 

\begin{equation}
    \text{Annualized Return} = \left(\left(1+\text{Return}\right)^{\text{n}/\text{N}}-1\right)\times 100  \label{annualized_return}
\end{equation}

Annualized rate of return allows comparison of different portfolios, but it doesn't take into account the risk related to the investment. Sharpe ratio, introduced by William Sharpe in 1966, is a risk-adjusted performance metric, and it shows the excess return of the portfolio per unit of risk. Equation \ref{sharpe} provides the overall formula of Sharpe Ratio, where \text{RF} is the risk-free rate. Therefore, the numerator represents the average excess return of the portfolio with respect to a risk-free asset, and the denominator is the standard deviation of the excess return of the portfolio. In this paper, we use risk-free rate is equal to zero for simplicity. 

\begin{equation}
    \text{Sharpe Ratio} = \frac{\mathbb{E}\left[\text{Return} - \text{RF}\right]}{\sigma} \label{sharpe}
\end{equation}
To compare the performance of the portfolios with different duration, we calculate the annualized Sharpe Ratio in Equation \ref{annualized_sharpe}. Annualized Sharpe ratio is calculated as the multiplication of Sharpe ratio from Equation \ref{sharpe} and the square-root of the period frequency within a year. 
\begin{equation}
    \text{Annualized Sharpe Ratio} = \text{Sharpe Ratio}\times \sqrt{n} \label{annualized_sharpe}
\end{equation}
Calmar ratio is another risk-adjusted performance metric that we considered in this paper. Instead of the standard deviation of excess returns, Calmar ratio uses maximum drawdown as the risk measure. Maximum drawdown is an indicatior of the largest loss that an investor can suffer. It is calculated as the maximum loss incurred by investing at the peak price and selling at the trough price. Equation \ref{calmar} show that Calmar ratio is simply formulated as the annualized return from Equation \ref{annualized_return} over the maximum drawdown. 

\begin{equation}
    \text{Calmar ratio} = \frac{\text{Annualized Return}}{\text{Maximum Drawdown}} \label{calmar}
\end{equation}

When evaluating the models in the following sections, we seek for attaining the maximum values for these three performance measures. If each performance indicates a different model or different lookback window, we then select the model and lookback window providing a balance between performance metrics. 

\subsection{Near-Term Results}
We predict one month ahead and three months ahead sector rankings to show how our model performs for near-term sector rankings.

\begin{table}[htbp]
\vskip\baselineskip 
\begin{center}
\begin{adjustwidth}{-1.2cm}{0cm}
\scalebox{0.7}{
\begin{tabular}{|c| l|| c c c||c c c|}\hline
\textbf{Lookback Window} & \textbf{Model} &\multicolumn{3}{c||}{\textbf{In-Sample Performance}}&\multicolumn{3}{c|}{ \textbf{Out-of-Sample Performance}}\\
\textbf{(Years)}& &{\small \textbf{Annualized Return}}&{\small \textbf{Sharpe Ratio}}&{\small \textbf{Calmar Ratio}}&{\small \textbf{Annualized Return}}&{\small \textbf{Sharpe Ratio}}&{\small \textbf{Calmar Ratio}}\\\hline
\rowcolor{blue!15}
\multicolumn{2}{|c||}{\textbf{Benchmark}} & 8.81\% & 0.660 & 0.175 & 13.60\% & 1.200 & 0.990  \\ \hline 
\multirow{4}{*}[-0.4ex]{{0.5}} & Ridge & 13.19\% & 0.858 & 0.318 & 10.22\% &0.953 & 0.633 \\ 
&LSTM & 17.70\% & 0.898 & 0.479 & 10.42\%  &1.212& 0.677 \\  
&GRU & 15.73\% & 0.934 & 0.417 & 10.38\% &1.148 & 0.643\\ 
&ESN & 17.89\% & 1.239 & 0.474 & 13.70\% &1.226 & 1.127 \\ 
\hline
\multirow{4}{*}[-0.4ex]{{1}} & Ridge & 13.32\% & 1.224 & 0.365 & 12.88\% &0.984 & 1.220 \\ 
&LSTM & 20.20\% & 1.271 & 0.602 & 13.96\%  &1.384& 1.386 \\  
&GRU & 17.05\% & 1.080 & 0.470 & 11.75\% &1.213 &0.773 \\ 
&ESN & 21.53\% & 1.034 & 0.689 & 12.05\% &1.457 & 0.809 \\ 
\hline
\multirow{4}{*}[-0.4ex]{{1.5}} & Ridge & 14.68\% & 0.960 & 0.431 & 10.34\% &1.074 & 0.796 \\ 
&LSTM & 19.66\% & 1.198 & 0.567 & 14.08\%  &1.332& 1.063 \\  
&GRU & 14.97\% & 0.840 & 0.343 & 9.55\% &1.091 & 0.522 \\ 
&ESN & 22.38\% & 1.286 & 0.757 & 15.46\% &1.517 & 1.028 \\ 
\hline
\multirow{4}{*}[-0.4ex]{{2}} & Ridge & 16.06\% & 1.126 & 0.450 & 13.02\% &1.169 & 0.970 \\ 
&LSTM & 20.98\% & 1.017 & 0.611 & 12.74\%  &1.381& 0.652 \\  
&GRU & 17.34\% & 1.048 & 0.486 & 11.70\% &1.194 & 0.813\\ 
&ESN & 24.25\% & 1.330 & 0.781 & 16.10\% &1.639 & 1.070 \\ 
\hline
\multirow{4}{*}[-0.4ex]{{2.5}} & Ridge & 16.56\% & 1.040 & 0.483 & 11.52\% &1.218 & 0.903 \\ 
&LSTM & 20.28\% & 1.111 & 0.658 & 13.04\%  & 1.411& 0.984 \\  
&GRU & 17.19\% & 0.965 & 0.401 & 10.26\% &1.210 & 0.784\\ 
&ESN & 24.72\% & 1.072 & 0.794 & 12.86\% &1.643 & 0.789 \\ 
\hline
\multirow{4}{*}[-0.4ex]{{3}} & Ridge & 16.93\% & 1.235 & 0.498 & 14.15\% &1.224 & 1.219 \\ 
&LSTM & 19.24\% & \textbf{1.466} & 0.610 & \textbf{17.83\%}  &1.342& \textbf{1.601} \\  
&GRU & 10.41\% & 0.931 & 0.193 & 10.49\% &0.775 & 0.602 \\ 
&ESN & \textbf{25.08\%} & 1.022 & \textbf{0.862} & 12.70\% &\textbf{1.702} & 0.777 \\ 
\hline
\end{tabular}}
\end{adjustwidth}
\end{center}
\caption{Next Month Prediction Results - Performance Measures}
\label{table:performance_1}
\end{table}
\FloatBarrier

Table \ref{table:performance_1} provides the performance of the portfolios formed based on different models and different lookback windows and compares these measures with that of benchmark portfolio for both training data and testing data. Results show that all models form superior portfolios in comparison to the benchmark portfolio under three performance metrics. For each lookback window, \texttt{ESN} provides the best annualized return and \texttt{LSTM} follows \texttt{ESN}. There is a similar pattern for annualized Calmar ratio. However, there is no clear pattern for annualized Sharpe ratio. For half of the lookback windows, \texttt{LSTM} performs the best, and \texttt{ESN} performs the best for other lookback horizons. Out-of-sample results show that \texttt{ESN} and \texttt{LSTM} based portfolios perform superior most of the time in terms of annualized returns. Similarly, annualized Sharpe ratios for these two models are superior to that of benchmark portfolio for each lookback window. \texttt{ESN} performs better than \texttt{LSTM} regarding annualized Sharpe ratio. There is no clear pattern for annualized Calmar ratio for the testing horizon.

\begin{figure}[!htbp]
\captionsetup{font=scriptsize,labelfont=scriptsize}
   \begin{minipage}{0.5\textwidth}
   \hspace*{-1.5cm}
     \centering
     \includegraphics[scale = 0.35]{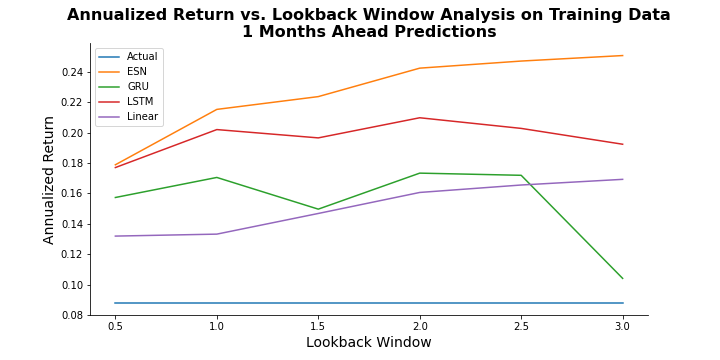}
   \end{minipage}\hfill
   \begin{minipage}{0.5\textwidth}
   \hspace*{0.5cm}
     \centering
     \includegraphics[scale = 0.35]{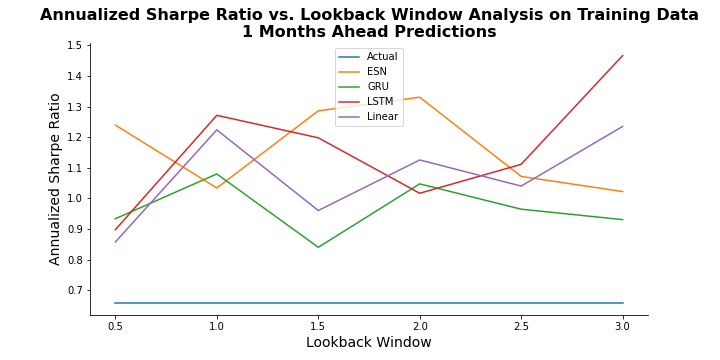}
   \end{minipage}\hfill
   \begin{minipage}{\textwidth}
     \centering
     \includegraphics[scale = 0.35]{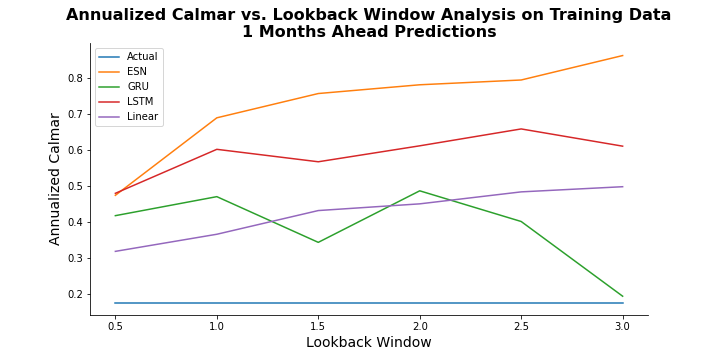}
   \end{minipage}
 \caption{Comparison of Different Lookback Windows and Different Models for Next Month Prediction} \label{fig:performance_training_1}
\end{figure}
\FloatBarrier

Figure \ref{fig:performance_training_1} illustrates how performance of different models change based on different lookback windows. Here, lookback window equals to 0.5 means that we use the last six months values of macroeconomic variables to predict a month ahead sector rankings. Plots show that portfolios formed based on all four models perform better than our benchmark portfolio during training horizon. Our results show that for annualized return and annualized Calmar ratio, \texttt{ESN} model creates a portfolio that is superior to other portfolios for all lookback windows, and \texttt{LSTM} follows \texttt{ESN}. There is no generic pattern for each model, but in general the annualized returns and annualized Calmar ratios increase as we use more historical data for predictions. For annualized Sharpe ratio, there is no strong correlation between lookback window and model performance. The best annualized Sharpe ratio is attained with three years historical data and \texttt{LSTM} model. Based on the performance measure criteria chosen, different models and different lookback windows are chosen for the final prediction model. If our priority is the annualized return or annualized Calmar ratio, then \texttt{ESN} with three years of historical data performs the best despite \texttt{LSTM} with three years of historical data is chosen for annualized Sharpe ratio. However, \texttt{ESN} with three years data do not give a nice annualized Sharpe ratio, and \texttt{LSTM} is not the best choice for the other performance metrics.

\begin{figure}[!htbp]
\captionsetup{font=scriptsize,labelfont=scriptsize}
   \begin{minipage}{0.5\textwidth}
   \hspace*{-1.5cm}
     \centering
     \includegraphics[scale = 0.35]{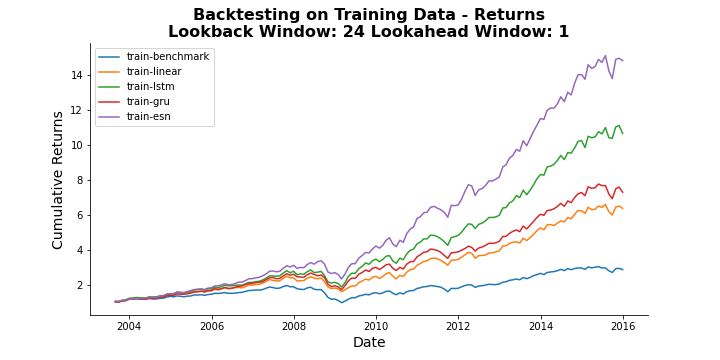}
   \end{minipage}\hfill
   \begin{minipage}{0.5\textwidth}
   \hspace*{0.5cm}
     \centering
     \includegraphics[scale = 0.35]{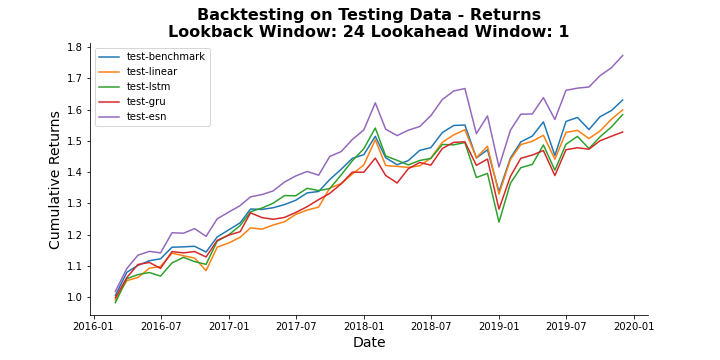}
   \end{minipage}
 \caption{Backtesting Performance for Next Month Prediction Models (Two Years Lookback Window)} \label{fig:backtesting_best_1}
\end{figure}
\FloatBarrier

To obtain nice results for each performance metric without sacrificing, we check the second best models for each performance metric. We observe that \texttt{ESN} with two years of historical data provides satisfactory performance during training horizon. Figure \ref{fig:backtesting_best_1} illustrates the backtesting results for different models with two years of historical data. All models exhibit superior performance in terms of cumulative returns during training horizon. \texttt{ESN} performs the best as expected. During the testing horizon, the performance of \texttt{ESN} is superior to the benchmark portfolio and other models. However, other models mostly underperform benchmark portfolio during testing period. There could be potential reasons for the changing performance during testing horizon. First, the benchmark portfolio performs better during testing period in comparison to training period. A thorough hyperparameter tuning could enhance the performance of other models during the testing period. Appendix \ref{backtesting_one_month} provides the performance metric plots for testing period and backtesting plots based on other lookback windows.

\begin{table}[htbp]
\vskip\baselineskip 
\begin{center}
\begin{adjustwidth}{-1.2cm}{0cm}
\scalebox{0.7}{
\begin{tabular}{|c| l|| c c c||c c c|}\hline
\textbf{Lookback Window} & \textbf{Model} &\multicolumn{3}{c||}{\textbf{In-Sample Performance}}&\multicolumn{3}{c|}{ \textbf{Out-of-Sample Performance}}\\
\textbf{(Years)}& &{\small \textbf{Annualized Return}}&{\small \textbf{Sharpe Ratio}}&{\small \textbf{Calmar Ratio}}&{\small \textbf{Annualized Return}}&{\small \textbf{Sharpe Ratio}}&{\small \textbf{Calmar Ratio}}\\\hline
\rowcolor{blue!15}
\multicolumn{2}{|c||}{\textbf{Benchmark}} & 7.93\% & 0.589 & 0.157 & 13.60\% & 1.200 & 0.990  \\ \hline 
\multirow{4}{*}[-0.4ex]{{1}} & Ridge & 11.06\% & 1.060 & 0.268 & 12.01\% &0.826 & 1.057 \\ 
&LSTM & 17.70\% & 1.051 & 0.504 & 12.93\%  &1.205& 0.837 \\  
&GRU & 7.11\% & 1.163 & 0.127 & 12.86\% &0.546 &1.087\\ 
&ESN & 18.95\% & 1.285 & 0.533 & 13.90\% &1.253 & 1.272 \\ 
\hline
\multirow{4}{*}[-0.4ex]{{2}} & Ridge & 15.48\% & 1.138 & 0.413 & 12.35\% &1.094 & 1.125 \\ 
&LSTM & 21.16\% & 1.417 & 0.643 & 16.59\%  &1.418& 1.268 \\  
&GRU & 11.92\% & 1.091 & 0.246 & 11.51\% &0.848 &0.836 \\ 
&ESN & 23.05\% & 1.397 & 0.738 & 16.02\% &1.534 & 1.450 \\ 
\hline
\multirow{4}{*}[-0.4ex]{{3}} & Ridge & 17.40\% & 1.172 & 0.508 & 13.08\% &1.212 & 1.501 \\ 
&LSTM & 19.90\% & 1.301 & 0.571 & 15.38\%  &1.321& 1.389 \\  
&GRU & 13.45\% & 1.322 & 0.344 & 15.51\% &0.960 & 1.169 \\ 
&ESN & 24.93\% & 1.527 & 0.768 & 16.46\% &1.635 & \textbf{1.958} \\ 
\hline
\multirow{4}{*}[-0.4ex]{{4}} & Ridge & 16.99\% & 1.111 & 0.496 & 11.65\% &1.192 & 1.149 \\ 
&LSTM & 20.23\% & 1.177 & 0.578 & 15.44\%  &1.321& 1.008 \\  
&GRU & 10.02\% & 0.923 & 0.218 & 11.23\% &0.696 & 0.614\\ 
&ESN & \textbf{27.22\%} & \textbf{1.574} & \textbf{0.894} & \textbf{17.58\%} &\textbf{1.746} & 1.575 \\ 
\hline
\end{tabular}}
\end{adjustwidth}
\end{center}
\caption{Three Months Ahead Prediction Results - Performance Measures}
\label{table:performance_3}
\end{table}
\FloatBarrier

In Table \ref{table:performance_3}, performance of different models based on different lookback windows are provided for three months ahead predictions. The results show that annualized returns obtained during training period are higher than that of benchmark portfolio for all models. As in Table \ref{table:performance_1}, portfolios based on \texttt{ESN} models provide the highest annualized returns during training period. The pattern is similar both for annualized Sharpe ratio and annualized Calmar ratio during training period. During the testing period, \texttt{ESN} and \texttt{LSTM} models attain better annualized returns compared to the benchmark and other two models. The best performance in terms of annualized Sharpe ratio and annualized Calmar ratio is attained also by \texttt{ESN} and \texttt{LSTM} models.

\begin{figure}[!htbp]
\captionsetup{font=scriptsize,labelfont=scriptsize}
   \begin{minipage}{0.5\textwidth}
   \hspace*{-1.5cm}
     \centering
     \includegraphics[scale = 0.35]{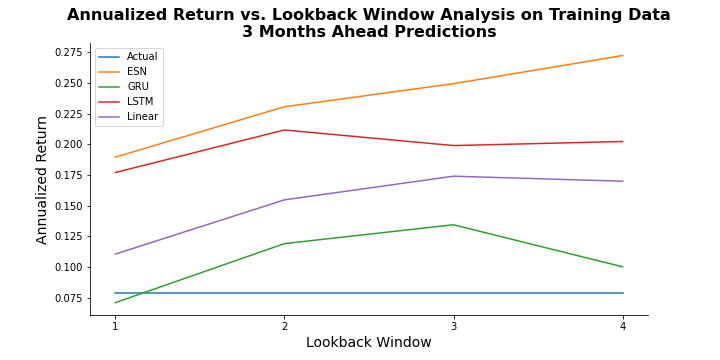}
   \end{minipage}\hfill
   \begin{minipage}{0.5\textwidth}
   \hspace*{0.5cm}
     \centering
     \includegraphics[scale = 0.35]{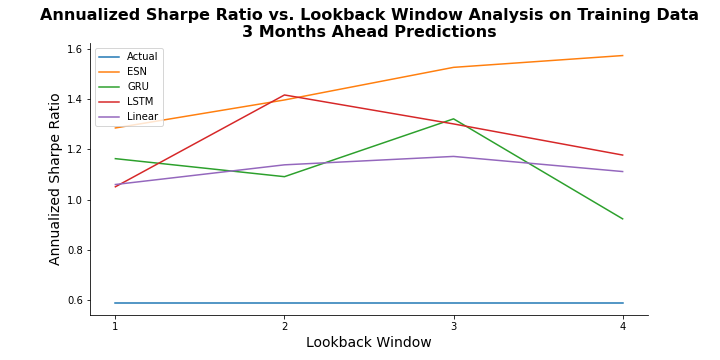}
   \end{minipage}\hfill
   \begin{minipage}{\textwidth}
     \centering
     \includegraphics[scale = 0.35]{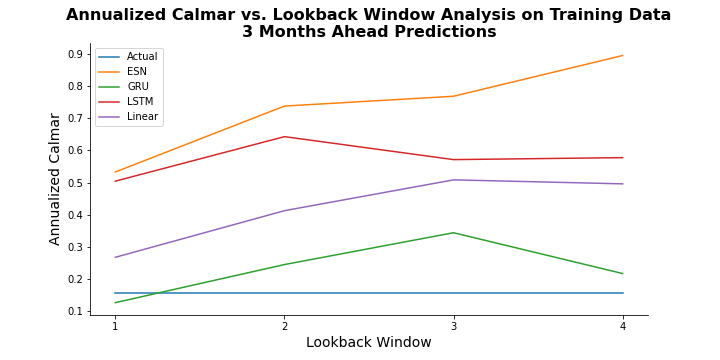}
   \end{minipage}
 \caption{Comparison of Different Lookback Windows and Different Models for Three Months Ahead Prediction} \label{fig:performance_training_3}
\end{figure}
\FloatBarrier
Figure \ref{fig:performance_training_3} illustrates the behavior of different models with different lookback windows for three months ahead predictions. For each performance metric plot, it is seen that \texttt{ESN} model performance increases as more historical data included in the model. For \texttt{LSTM} model, the performance increases from one year to two years, but then it decreases as we include more historical data. Similarly, \texttt{GRU} and ridge model performances increase from one year to three years but then decreases. For all performance metrics, \texttt{ESN} with four years of historical data provides the best performance. 

\begin{figure}[!htbp]
\captionsetup{font=scriptsize,labelfont=scriptsize}
   \begin{minipage}{0.5\textwidth}
   \hspace*{-1.5cm}
     \centering
     \includegraphics[scale = 0.35]{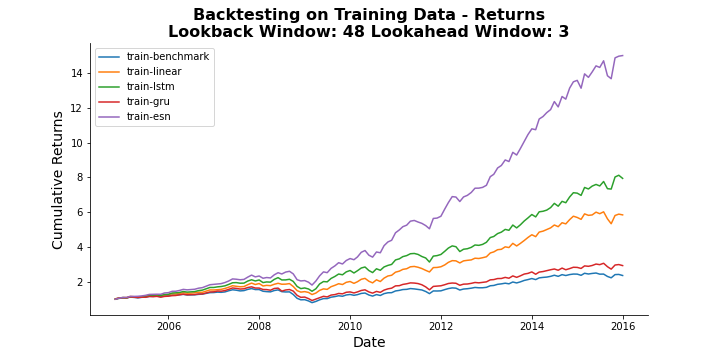}
   \end{minipage}\hfill
   \begin{minipage}{0.5\textwidth}
   \hspace*{0.5cm}
     \centering
     \includegraphics[scale = 0.35]{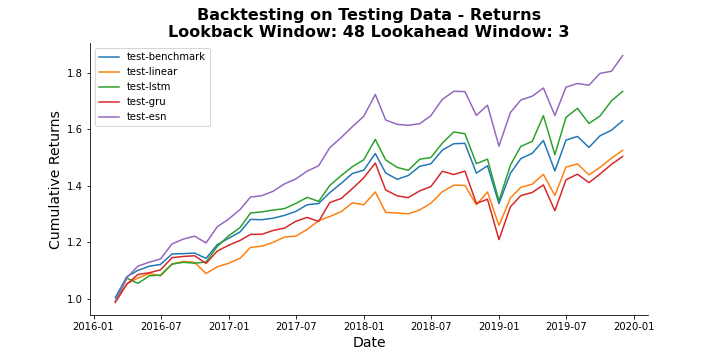}
   \end{minipage}
 \caption{Backtesting Performance for Three Months Ahead Prediction Models (Four Years Lookback Window)} \label{fig:backtesting_best_3}
\end{figure}
\FloatBarrier

Figure \ref{fig:backtesting_best_3} shows the backtesting results for the models that use four years of historical data. During training period, portfolios formed by all models outperform the benchmark portfolio. \texttt{ESN} performs the best, and \texttt{LSTM} follows \texttt{ESN}. Testing period cumulative returns show that \texttt{ESN} ourperforms all the models and also benchmark period. \texttt{LSTM} model also provides better cumulative returns over testing period in comparison to benchmark portfolio. Appendix \ref{backtesting_three_months} consists of the performance metric plots for testing period and also the backtesting results for other lookback windows. 

\subsection{Medium-Term Results}
We predict six months ahead and one year ahead sector rankings to show how our methodology works for medium-term sector rankings.
\begin{table}[htbp]
\vskip\baselineskip 
\begin{center}
\begin{adjustwidth}{-1.2cm}{0cm}
\scalebox{0.7}{
\begin{tabular}{|c| l|| c c c||c c c|}\hline
\textbf{Lookback Window} & \textbf{Model} &\multicolumn{3}{c||}{\textbf{In-Sample Performance}}&\multicolumn{3}{c|}{ \textbf{Out-of-Sample Performance}}\\
\textbf{(Years)}& &{\small \textbf{Annualized Return}}&{\small \textbf{Sharpe Ratio}}&{\small \textbf{Calmar Ratio}}&{\small \textbf{Annualized Return}}&{\small \textbf{Sharpe Ratio}}&{\small \textbf{Calmar Ratio}}\\\hline
\rowcolor{blue!15}
\multicolumn{2}{|c||}{\textbf{Benchmark}} & 7.24\% & 0.544 & 0.144 & 13.60\% & 1.200 & 0.990  \\ \hline 
\multirow{4}{*}[-0.4ex]{{1}} & Ridge & 12.95\% & 0.760 & 0.302 & 8.17\% &0.885 & 0.562 \\ 
&LSTM & 19.19\% & 1.460 & 0.577 & 16.24\%  &1.256& 1.386 \\  
&GRU & 9.05\% & 1.172 & 0.159 & 13.26\% &0.653 &1.090\\ 
&ESN & 17.88\% & 1.434& 0.482 & \textbf{16.73\%} &1.191 & 1.371 \\ 
\hline
\multirow{4}{*}[-0.4ex]{{2}} & Ridge & 13.59\% & 1.083 & 0.361 & 11.60\% &0.984 & 1.117 \\ 
&LSTM & 18.42\% & 1.415 & 0.514 & 16.50\%  &1.251& 1.247 \\  
&GRU & 9.39\% & 1.183 & 0.201 & 12.61\% &0.684 &1.500 \\ 
&ESN & 20.61\% & \textbf{1.477} & 0.584 & 16.01\% &1.355 & 1.411 \\ 
\hline
\multirow{4}{*}[-0.4ex]{{3}} & Ridge & 15.29\% & 1.352 & 0.402 & 14.84\% &1.077 & 1.406 \\ 
&LSTM & 18.12\% & 1.139 & 0.548 & 13.98\%  &1.244& 0.908 \\  
&GRU & 11.84\% & 1.136 & 0.239 & 13.31\% &0.867 & 0.906 \\ 
&ESN & 24.76\% & 1.393 & \textbf{0.761} & 15.71\% &\textbf{1.622} & \textbf{1.611} \\ 
\hline
\multirow{4}{*}[-0.4ex]{{4}} & Ridge & 16.18\% & 1.214 & 0.477 & 13.57\% &1.138 & 1.281 \\ 
&LSTM & 18.98\% & 1.304 & 0.565 & 15.95\%  &1.267& 1.065 \\  
&GRU & 13.18\% & 1.322 & 0.289 & 16.34\% &0.931 & 1.232\\ 
&ESN & \textbf{25.34\%} & 1.252 & 0.732 & 13.92\% &1.606 & 1.379 \\ 
\hline
\end{tabular}}
\end{adjustwidth}
\end{center}
\caption{Six Months Ahead Prediction Results - Performance Measures}
\label{table:performance_6}
\end{table}
\FloatBarrier
Table \ref{table:performance_6} shows the performance of different models based on different lookback windows for six month ahead predictions. The results are consistent with near-term results because \texttt{ESN} and \texttt{LSTM} models outperform other models and benchmark portfolio in terms of annualized rate of return and annualized Calmar ratio during training period. For annualized Sharpe ratio, \texttt{ESN} and \texttt{LSTM} models are better in general during training period, but other models obtain satisfactory outcomes, as well. Mostly, \texttt{ESN} model performs the best during testing period for annualized Sharpe ratio and annualized Calmar ratio. Consistently, \texttt{ESN} and \texttt{LSTM} obtain higher annualized return during testing period. However, \texttt{GRU} and ridge regression models obtain good returns, as well. 
\begin{figure}[!htbp]
\captionsetup{font=scriptsize,labelfont=scriptsize}
   \begin{minipage}{0.5\textwidth}
   \hspace*{-1.5cm}
     \centering
     \includegraphics[scale = 0.35]{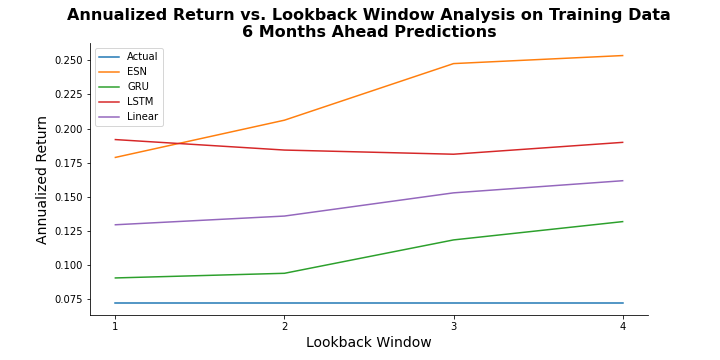}
   \end{minipage}\hfill
   \begin{minipage}{0.5\textwidth}
   \hspace*{0.5cm}
     \centering
     \includegraphics[scale = 0.35]{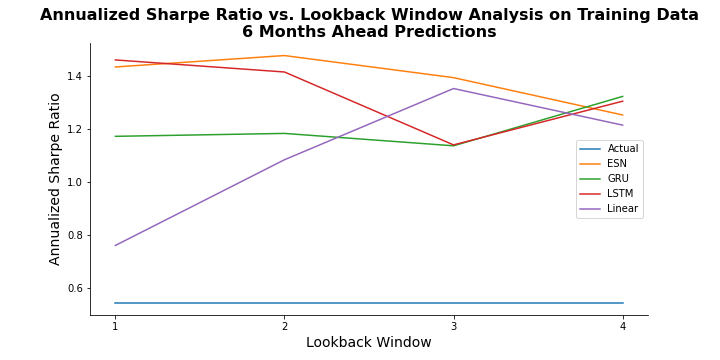}
   \end{minipage}\hfill
   \begin{minipage}{\textwidth}
     \centering
     \includegraphics[scale = 0.35]{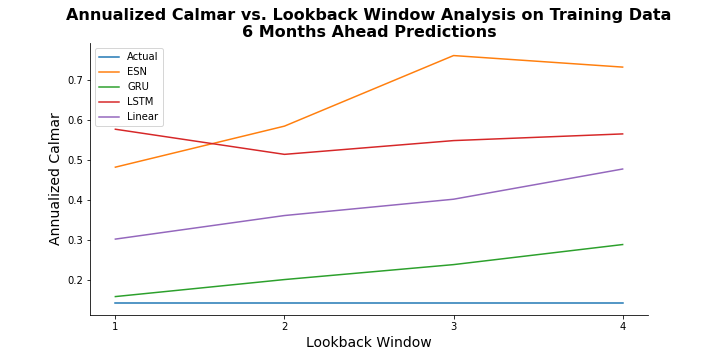}
   \end{minipage}
 \caption{Comparison of Different Lookback Windows and Different Models for Six Months Ahead Prediction} \label{fig:performance_training_6}
\end{figure}
\FloatBarrier
Figure \ref{fig:performance_training_6} indicates that \texttt{ESN} model performs the best in general for all metric during training period. The best annualized return is obtained for \texttt{ESN} with four years historical data. The annualized return tend to increase with an increase in lookback window for \texttt{ESN}, \texttt{GRU}, and ridge regression models. There is no clear pattern for annualized Sharpe ratio. The highest values are obtained for \texttt{ESN} with two years of historical data, and then \texttt{LSTM} with one year of historical data. Annualized Calmar ratio increases with longer historical data for \texttt{GRU} and ridge regression models, but the relationship is not linear for other models. The best annualized Calmar ratio is obtained for \texttt{ESN} with three years of historical data. 

\begin{figure}[!htbp]
\captionsetup{font=scriptsize,labelfont=scriptsize}
   \begin{minipage}{0.5\textwidth}
   \hspace*{-1.5cm}
     \centering
     \includegraphics[scale = 0.35]{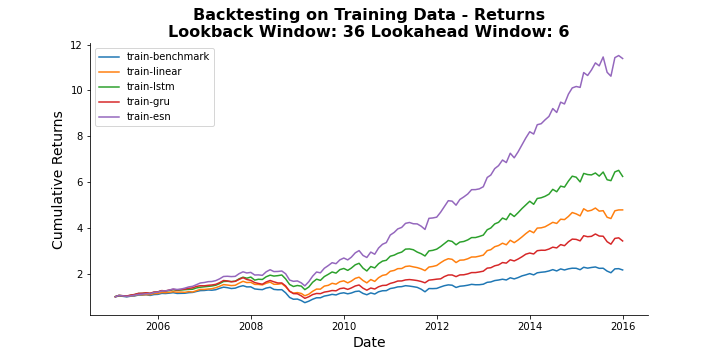}
   \end{minipage}\hfill
   \begin{minipage}{0.5\textwidth}
   \hspace*{0.5cm}
     \centering
     \includegraphics[scale = 0.35]{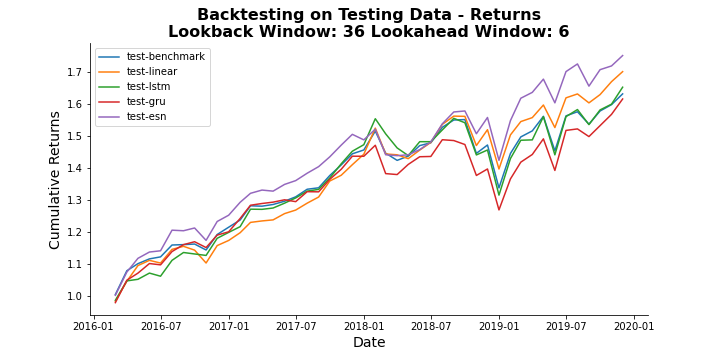}
   \end{minipage}
 \caption{Backtesting Performance for Six Months Ahead Prediction Models (Three Years Lookback Window)} \label{fig:backtesting_best_6}
\end{figure}
\FloatBarrier

As seen in Figure \ref{fig:performance_training_6}, \texttt{ESN} is the superior model during training period for six month ahead predictions but different lookback windows obtain the highest values for each performance metric. In order to balance the performance, we pick \texttt{ESN} model with three years of historical data. Figure \ref{fig:backtesting_best_6} shows the backtesting results for different models with three years of historical data. Consistent with near-term prediction results, all portfolios built by different models perform much better than the benchmark portfolio during training period, and the portfolio which is built based on \texttt{ESN} model obtains highest values. During testing period, \texttt{ESN} performs the best except the first half of 2018 where \texttt{LSTM} slightly outperforms. All models except \texttt{GRU} outperforms the benchmark portfolio within testing period. Appendix \ref{backtesting_six_months} provides the performance metric plots during testing period, and the backtesting results for other lookback windows for six months ahead predictions.

\begin{table}[htbp]
\vskip\baselineskip 
\begin{center}
\begin{adjustwidth}{-1.2cm}{0cm}
\scalebox{0.7}{
\begin{tabular}{|c| l|| c c c||c c c|}\hline
\textbf{Lookback Window} & \textbf{Model} &\multicolumn{3}{c||}{\textbf{In-Sample Performance}}&\multicolumn{3}{c|}{ \textbf{Out-of-Sample Performance}}\\
\textbf{(Years)}& &{\small \textbf{Annualized Return}}&{\small \textbf{Sharpe Ratio}}&{\small \textbf{Calmar Ratio}}&{\small \textbf{Annualized Return}}&{\small \textbf{Sharpe Ratio}}&{\small \textbf{Calmar Ratio}}\\\hline
\rowcolor{blue!15}
\multicolumn{2}{|c||}{\textbf{Benchmark}} & 7.40\% & 0.547 & 0.147 & 13.60\% & 1.200 & 0.990  \\ \hline 
\multirow{4}{*}[-0.4ex]{{1}} & Ridge & 12.26\% & 1.136 & 0.288 & 12.54\% &0.827 & 0.924 \\ 
&LSTM & 16.80\% & 1.386 & 0.498 & \textbf{16.46\%}  &1.166& 1.350 \\  
&GRU & 12.78\% & 1.051 & 0.238 & 12.21\% &0.830 &0.869\\ 
&ESN & 20.94\% & \textbf{1.530}& 0.617 & 15.35\% &1.389 & 1.554 \\ 
\hline
\multirow{4}{*}[-0.4ex]{{2}} & Ridge & 13.95\% & 1.342 & 0.346 & 15.54\% &0.956 & 1.351 \\ 
&LSTM & 21.18\% & 1.206 & 0.585 & 13.75\%  &1.378& 1.100 \\  
&GRU & 11.26\% & 1.173 & 0.215 & 13.45\% &0.743 &1.125 \\ 
&ESN & 20.81\% & 1.346 & 0.589 & 15.13\% &1.391 & 1.397 \\ 
\hline
\multirow{4}{*}[-0.4ex]{{3}} & Ridge & 16.42\% & 1.289 & 0.444 & 14.00\% &1.129 & \textbf{1.614} \\ 
&LSTM & 19.54\% & 1.159 & 0.566 & 12.11\%  &1.308& 1.054 \\  
&GRU & 13.42\% & 1.423 & 0.272 & 15.92\% &0.878 & 1.440 \\ 
&ESN & 23.38\% & 1.091 & 0.679 & 12.50\% &1.528 & 0.959\\ 
\hline
\multirow{4}{*}[-0.4ex]{{4}} & Ridge & 17.00\% & 1.195 & 0.473 & 13.95\% &1.186 & 1.133 \\ 
&LSTM & 20.62\% & 1.269 & 0.604 & 14.70\%  &1.372& 1.111 \\  
&GRU & 10.83\% & 1.186 & 0.218 & 14.94\% &0.727 & 1.189\\ 
&ESN & \textbf{25.08\%} & 1.206 & \textbf{0.727} & 15.36\% &\textbf{1.569} & 1.207 \\ 
\hline
\end{tabular}}
\end{adjustwidth}
\end{center}
\caption{One Year Ahead Prediction Results - Performance Measures}
\label{table:performance_12}
\end{table}
\FloatBarrier

Table \ref{table:performance_12} shows the performance of different models based on different amount of historical data for one year ahead predictions. In consistent to other prediction models, \texttt{ESN} and \texttt{LSTM} models outperform other models and benchmark portfolio in terms of annualized return and annualized Calmar ratio during training horizon. Despite the pattern is similar for annualized Sharpe ratio, \texttt{GRU} and ridge regression models also perform well for two years and three years of historical data. Testing period results indicate that \texttt{ESN} model is superior over other models if annualized Sharpe ratio is chosen as a performance metric. \texttt{ESN} model provides admissable annualized returns and annualized Calmar ratio, but the other models also outperform \texttt{ESN} model for different lookback windows. 

\begin{figure}[!htbp]
\captionsetup{font=scriptsize,labelfont=scriptsize}
   \begin{minipage}{0.5\textwidth}
   \hspace*{-1.5cm}
     \centering
     \includegraphics[scale = 0.35]{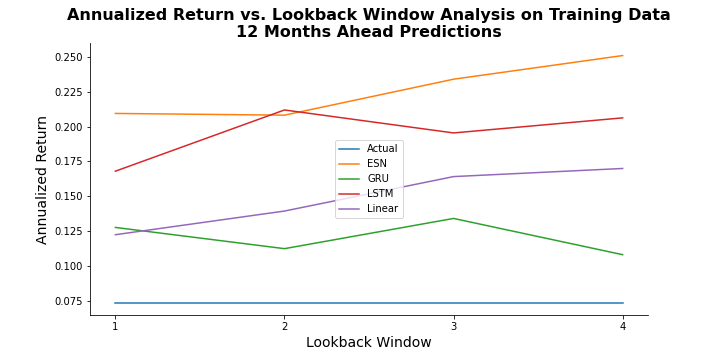}
   \end{minipage}\hfill
   \begin{minipage}{0.5\textwidth}
   \hspace*{0.5cm}
     \centering
     \includegraphics[scale = 0.35]{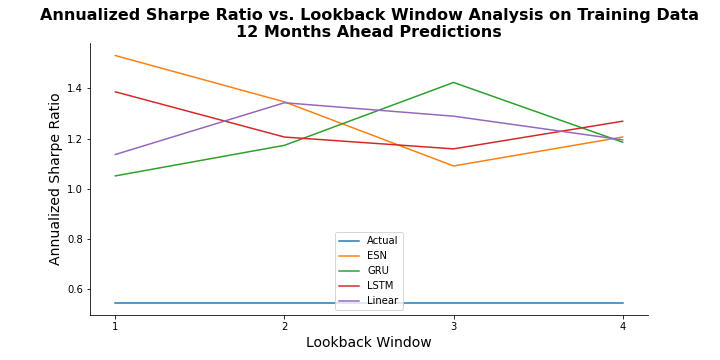}
   \end{minipage}\hfill
   \begin{minipage}{\textwidth}
     \centering
     \includegraphics[scale = 0.35]{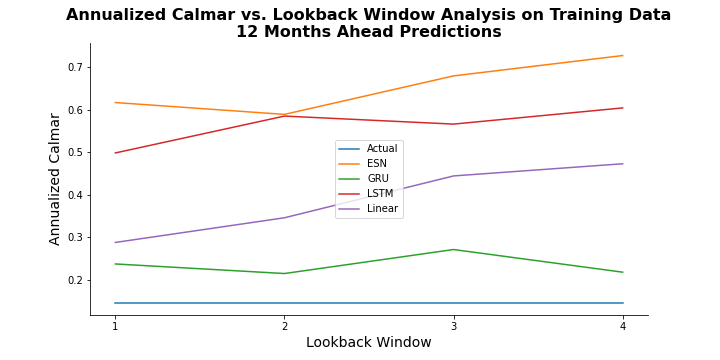}
   \end{minipage}
 \caption{Comparison of Different Lookback Windows and Different Models for One Year Ahead Prediction} \label{fig:performance_training_12}
\end{figure}
\FloatBarrier
Figure \ref{fig:performance_training_12} illustrates the relationship between different performance metrics and lookback window length for different models. Plots show that there is a positive relationship between annualized return and lookback window for \texttt{ESN} and ridge regression models during training period. Moreover, \texttt{ESN} achieves the best values for these performance metrics. Annualized Sharpe ratio decreases until three years of lookback window, and then it starts increasing for \texttt{ESN} and \texttt{LSTM} models. None of the models constantly outperforms the other models when annualized Sharpe ratio is the performance metric. The best value of annualized Sharpe ratio is attained with \texttt{ESN} model and one year of historical data. 
\begin{figure}[!htbp]
\captionsetup{font=scriptsize,labelfont=scriptsize}
   \begin{minipage}{0.5\textwidth}
   \hspace*{-1.5cm}
     \centering
     \includegraphics[scale = 0.35]{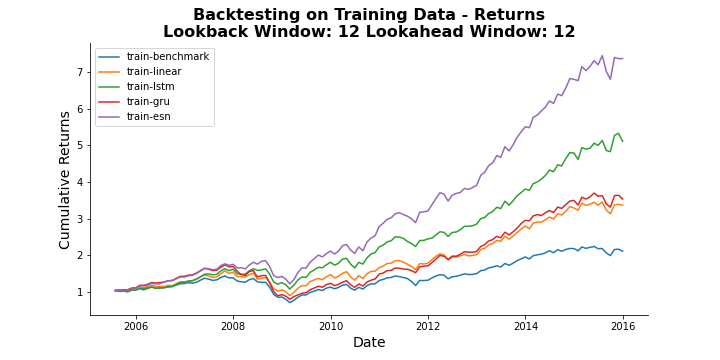}
   \end{minipage}\hfill
   \begin{minipage}{0.5\textwidth}
   \hspace*{0.5cm}
     \centering
     \includegraphics[scale = 0.35]{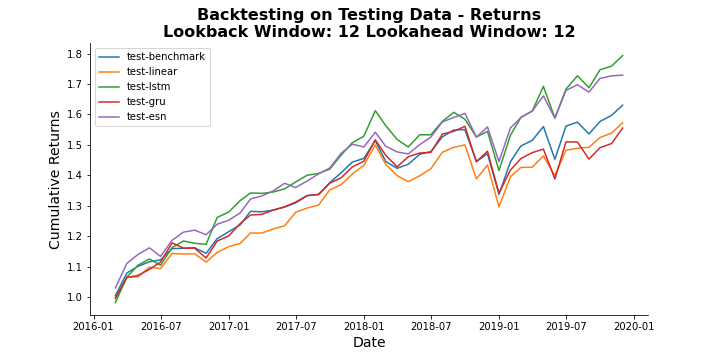}
   \end{minipage}
 \caption{Backtesting Performance for One Year Ahead Prediction Models (One Year Lookback Window)} \label{fig:backtesting_best_12}
\end{figure}
\FloatBarrier
Figure \ref{fig:performance_training_12} shows that \texttt{ESN} is superior to others when the performance metric is annualized return or annualized Calmar ratio, but there is no strong conclusion for annualized Sharpe ratio. Although the best values of annualized return and annualized Calmar ratio are both attained using \texttt{ESN} with four years of historical data, the value of annualized Sharpe ratio with the same setting is much lower than that of other models and other lookback windows. On the other hand, the best annualized Sharpe ratio is attained with \texttt{ESN} and one year of historical data. We select \texttt{ESN} with one year of historical data because the third best values for annualized return and annualized Calmar ratio are also obtained with \texttt{ESN} and one year lookback window. Figure \ref{fig:backtesting_best_12} illustrates the backtesting results for all models which make use of  one year of historical data. Training period results are in accordance with other prediction results. \texttt{ESN} model outperforms the other models, and all models perform better than equally weighted benchmark portfolio. \texttt{ESN} and \texttt{LSTM} models achieve similar cumulative returns over testing period, and they beat the performance of the benchmark portfolio significantly. Appendix \ref{backtesting_one_year} includes the backtesting results for other lookback windows together with the performance metric plots for testing period for one year ahead sector ranking predictions. 

\subsection{Long-Term Results}
Lastly, we observe the performance of our methodology in the long run by applying on two years ahead sector rankings prediction. We employ all models for different lookback windows ranging from one to five years. 

\begin{table}[htbp]
\vskip\baselineskip 
\begin{center}
\begin{adjustwidth}{-1.2cm}{0cm}
\scalebox{0.7}{
\begin{tabular}{|c| l|| c c c||c c c|}\hline
\textbf{Lookback Window} & \textbf{Model} &\multicolumn{3}{c||}{\textbf{In-Sample Performance}}&\multicolumn{3}{c|}{ \textbf{Out-of-Sample Performance}}\\
\textbf{(Years)}& &{\small \textbf{Annualized Return}}&{\small \textbf{Sharpe Ratio}}&{\small \textbf{Calmar Ratio}}&{\small \textbf{Annualized Return}}&{\small \textbf{Sharpe Ratio}}&{\small \textbf{Calmar Ratio}}\\\hline
\rowcolor{blue!15}
\multicolumn{2}{|c||}{\textbf{Benchmark}} & 5.36\% & 0.400 & 0.106 & 13.60\% & 1.200 & 0.990  \\ \hline 
\multirow{4}{*}[-0.4ex]{{1}} & Ridge & 12.73\% & 1.042 & 0.334 & 12.22\% &0.852 & 0.885 \\ 
&LSTM & 15.01\% & 1.219 & 0.369 & 12.43\%  &0.983& 1.095 \\  
&GRU & 12.11\% & 1.397 & 0.267 & 16.74\% &0.789 &1.484\\ 
&ESN & 18.52\% & 1.211& 0.529 & 14.52\% &1.148 & 1.314 \\ 
\hline
\multirow{4}{*}[-0.4ex]{{2}} & Ridge & 15.14\% & 1.122 & 0.420 & 13.08\% &1.010 & 1.137 \\ 
&LSTM & 18.56\% & 1.271 & 0.539 & 16.13\%  &1.150& 1.053 \\  
&GRU & 5.85\% & 1.079 & 0.110 & 13.20\% &0.420 &0.836 \\ 
&ESN & 20.51\% & 1.308 & 0.622 & 15.40\% &1.277 & 1.084 \\ 
\hline
\multirow{4}{*}[-0.4ex]{{3}} & Ridge & 15.37\% & 1.402 & 0.428 & 15.73\% &1.054 & 1.699\\ 
&LSTM & 19.18\% & 1.369 & 0.552 & 13.93\%  &1.216& 1.512 \\  
&GRU & 8.49\% & 0.991 & 0.157 & 12.02\% &0.578& 0.852 \\ 
&ESN & 23.06\% & 1.119 & 0.681 & 13.24\% &1.378 & 0.974\\ 
\hline
\multirow{4}{*}[-0.4ex]{{4}} & Ridge & 16.60\% & \textbf{1.568} & 0.467 & \textbf{18.72\%} &1.101 & \textbf{2.159}\\ 
&LSTM & 17.64\% & 1.451 & 0.516 & 15.51\%  &1.142& 1.417 \\  
&GRU & 13.34\% & 1.067 & 0.331 & 13.06\% &0.860 & 0.860\\ 
&ESN & 24.04\% & 1.192 & 0.751 & 14.27\% &1.432 & 1.028 \\ 
\hline
\multirow{4}{*}[-0.4ex]{{5}} & Ridge & 17.13\% & 1.179 & 0.472 & 14.02\% &1.144 & 1.220 \\ 
&LSTM & 18.42\% & 1.448 & 0.533 & 17.44\%  &1.126& 1.277 \\  
&GRU & 7.06\% & 1.091 & 0.124 & 12.54\% &0.493 & 0.964\\ 
&ESN & \textbf{24.70\%} & 1.185 & \textbf{0.786} & 15.16\% &\textbf{1.480} & 0.892\\ 
\hline
\end{tabular}}
\end{adjustwidth}
\end{center}
\caption{Two Years Ahead Prediction Results - Performance Measures}
\label{table:performance_24}
\end{table}
\FloatBarrier

Table \ref{table:performance_24} gives the overview of the performance of the models with different historical data span. During the training period, \texttt{ESN} achieves the highest annualized return and annualized Calmar ratio values. Despite \texttt{ESN} obtains desirable annualized Sharpe ratio values, other models may perform better than \texttt{ESN} for different lookback windows. In addition, \texttt{ESN} attains the highest annualized Sharpe ratios within testing period. During testing period, \texttt{ESN} performs moderate compared to other models for other metrics, but still outperforms the benchmark portfolio.
\begin{figure}[!htbp]
\captionsetup{font=scriptsize,labelfont=scriptsize}
   \begin{minipage}{0.5\textwidth}
   \hspace*{-1.5cm}
     \centering
     \includegraphics[scale = 0.35]{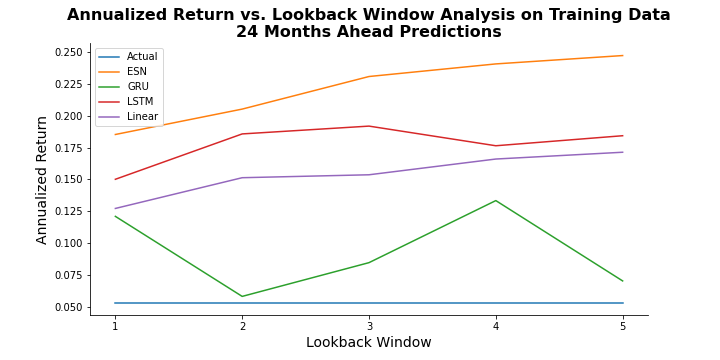}
   \end{minipage}\hfill
   \begin{minipage}{0.5\textwidth}
   \hspace*{0.5cm}
     \centering
     \includegraphics[scale = 0.35]{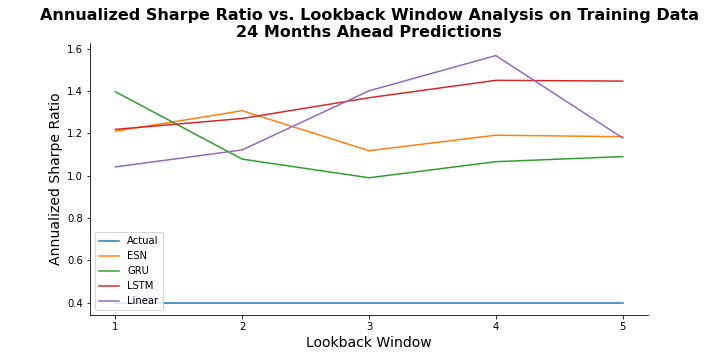}
   \end{minipage}\hfill
   \begin{minipage}{\textwidth}
     \centering
     \includegraphics[scale = 0.35]{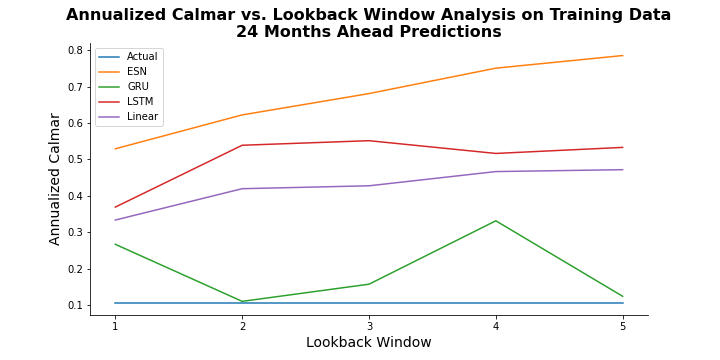}
   \end{minipage}
 \caption{Comparison of Different Lookback Windows and Different Models for Two Years Ahead Prediction} \label{fig:performance_training_24}
\end{figure}
\FloatBarrier
Figure \ref{fig:performance_training_24} shows that annualized return and annualized Calmar ratio follow similar patterns. The performance of \texttt{ESN} and ridge regression increases with an increase in the lookback window. \texttt{ESN} obtains highest values at each lookback window. The superiority of the models is not obvious from annualized Sharpe ratio plot. The plots indicate that \texttt{ESN} with five years of historical data attains the best annualized return and annualized Calmar ratio values. We select \texttt{ESN} with five years of historical data for illustrative purposes because it also obtains moderate annualized Sharpe ratio.
\begin{figure}[!htbp]
\captionsetup{font=scriptsize,labelfont=scriptsize}
   \begin{minipage}{0.5\textwidth}
   \hspace*{-1.5cm}
     \centering
     \includegraphics[scale = 0.35]{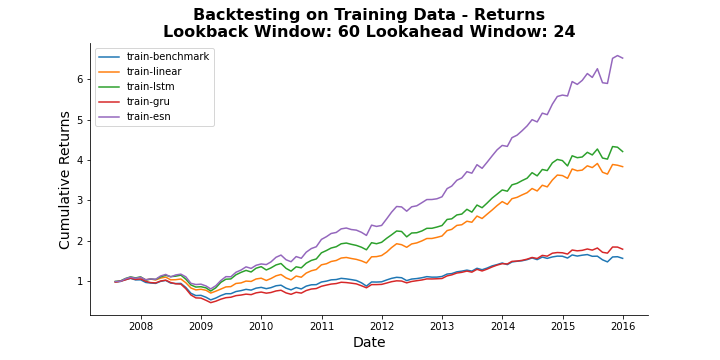}
   \end{minipage}\hfill
   \begin{minipage}{0.5\textwidth}
   \hspace*{0.5cm}
     \centering
     \includegraphics[scale = 0.35]{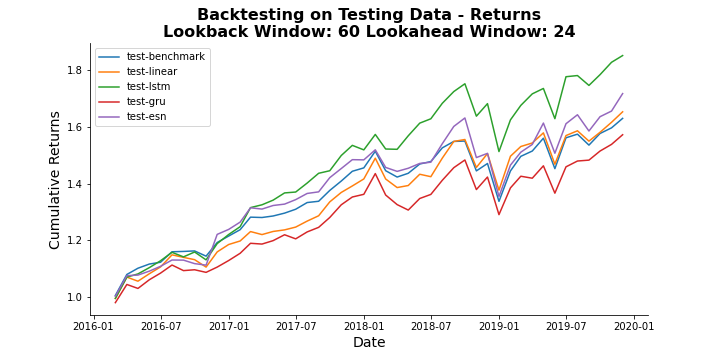}
   \end{minipage}
 \caption{Backtesting Performance for Two Years Ahead Prediction Models (Five Years Lookback Window)} \label{fig:backtesting_best_24}
\end{figure}
\FloatBarrier

Figure \ref{fig:backtesting_best_24} reveals the backtesting results of the models with the existence of five years historical data knowledge. Training period results are consistent with all other prediction horizon results. Although \texttt{ESN} outperforms the benchmark portfolio, \texttt{LSTM} performs better than \texttt{ESN} and other models. Appendix \ref{backtesting_two_years} consists of the remaining backtesting results and performance metric plots over testing period for two years ahead predictions. 

\subsection{Diagnostics}
In this section, we analyze the experimental results obtained by implementing the proposed methodology suggested in Section \ref{Methodology-MI}. 
\begin{itemize}

\item Based on our experiments, we observe that our methodology is able to beat the performance of the benchmark portfolio no matter how long ahead we predict the sector rankings. Furthermore, our strategy for picking the models and lookback windows based on the balanced performance metrics always outperforms the benchmark portfolio for all prediction periods. 
\item The best performance is usually attained when the lookback window takes the largest value in the experiment range.  

\item Annualized return and annualized Calmar plots tend to follow the similar patterns for both training and testing horizons. The behavior of \texttt{LSTM} and \texttt{ESN} models are similar with the change of lookback window. \texttt{GRU} and ridge regression models also follow similar path according to lookback window. Although the relationship between the performance of other models and lookback windows is not clear, \texttt{ESN} model exhibits an increasing annualized return and annualized Calmar ratio as the lookback window increases. 

\item For each prediction period, all models achieve better cumulative returns than benchmark portfolio over training horizon. Moreover, the cumulative returns of all models are sorted in descending order as \texttt{ESN}, \texttt{LSTM}, ridge regression, \texttt{GRU}, and the benchmark portfolio. During testing period, \texttt{ESN} and \texttt{LSTM} models almost always outperform the benchmark portfolio. Although \texttt{GRU} and ridge regression models also provide admissible results, their relative performance according to benchmark is not stable enough. 
\item In comparison to \texttt{LSTM} and \texttt{GRU}, \texttt{ESN} model is easier to implement and faster. \texttt{ESN} model outperforms all other models and benchmark portfolio during training horizon, and the outstanding performance of \texttt{ESN} model also continues over testing horizon most of the time. 

\item In this paper, we chose a certain set of hyperparameters based on our preliminary experiments, but integrating hyperparameter tuning techniques such Bayesian optimization into our methodology may boost up the performance of the models and hence the overall methodology. 

\end{itemize}

\section{Conclusion and Future Work} \label{Conclusion - MI}
In this paper, we develop a methodology for ranking sectors using a variety of common and sector-specific macroeconomic indicators. The methodology consists of two stages: prediction of sector index prices, and ranking sectors based on their predicted rate of returns. For each sector, we first apply \texttt{RFE} model to identify the most significant features, and then implement one of ridge regression, \texttt{LSTM}, \texttt{GRU}, and \texttt{ESN} models to predict the future sector index prices. We then rank the sectors based on their predicted rate of returns, and select top four ranked sectors for long-only strategy. We test the performance of our methodology over short-term, mid-term, and long-term. For each model, we also experiment with different lookback windows to show the robustness of the models based on the historical data used. Our numerical results show that the models generally performs better in the existence of longer lookback windows. We evaluate the performance of the models based on annualized return, annualized Sharpe ratio, and Calmar ratio. We pick the best performing model as the model satisfies admissible values for all performance measures. For each prediction horizon, \texttt{ESN} model is chosen based on this strategy. Our backtesting results indicate that our strategy that is built based on our methodology beats the performance of the benchmark portfolio consistently.

This paper consists of two consecutive stages where we first implement prediction models for each sector independently, and then rank the sectors based on their predicted returns. As a next step, we are going to develop a single step model that directly predicts the ranking of the sectors based on the market environment because employing prediction models separately for each sector increase the uncertainty of the overall performance. Our purpose is to implement learning-to-rank algorithms such as ListNet, BayesRank, and BoltzRank since they are specifically built for listwise ranking. We also plan to develop a methodology for obtaining sector-specific sentiment scores based on news articles, and integrate these sentiment scores into our current methodology as an input. Lastly, we build our methodology using public market information in this paper. For the future work, we aim at extending this methodology to private equity funds.

\clearpage

\printbibliography

\newpage
\appendix
\renewcommand{\thetable}{\Alph{section}\arabic{table}}
\addcontentsline{toc}{section}{Appendices}
\section*{Appendices}
\section{Macroeconomic Variables} \label{macroeconomic}

\begin{table}[htbp]
\vskip\baselineskip 
\begin{center}
\scalebox{0.8}{
\begin{tabular}{|l|c|c|} \hline
\textbf{Variables}& \textbf{Source} & \textbf{Frequency} \\ \hline
    GDP &  \url{https://fred.stlouisfed.org} & Quarterly \\ \hline
    Unemployment Rate & \url{https://fred.stlouisfed.org} & Monthly  \\ \hline
    CPI &\url{https://fred.stlouisfed.org}& Monthly \\ \hline
    MORTGAGE30US & \url{https://fred.stlouisfed.org} & Weekly \\ \hline
    Effective Federal Funds Rate & \url{https://fred.stlouisfed.org} & Monthly \\ \hline
\end{tabular}}
\end{center}
\caption{Macroeconomic Variables Common for All Sectors}
\label{table:all_sectors}
\end{table}

\begin{table}[htbp]
\vskip\baselineskip 
\begin{center}
\scalebox{0.8}{
\begin{tabular}{|l|c|c|} \hline
\textbf{Variables}& \textbf{Source} & \textbf{Frequency} \\ \hline
    Life Expectancy & \url{https://www.macrotrends.net} & Annual \\ \hline
    Population & \url{https://www.macrotrends.net} & Annual  \\ \hline
    Birth Rate &\url{https://www.macrotrends.net}& Annual \\ \hline
    Death Rate & \url{https://www.macrotrends.net} & Annual \\ \hline
\end{tabular}}
\end{center}
\caption{Macroeconomic Variables Considered for Healthcare Sector}
\label{table:healthcare_sector}
\end{table}

\begin{table}[htbp]
\vskip\baselineskip 
\begin{center}
\scalebox{0.8}{
\begin{tabular}{|l|c|c|} \hline
\textbf{Variables}& \textbf{Source} & \textbf{Frequency} \\ \hline
    U.S. Inflation Rate &  \url{https://www.macrotrends.net} & Annual \\ \hline
    5 Year Forward Inflation Rate & \url{https://www.macrotrends.net} & Annual  \\ \hline
    LIBOR Rate & \url{https://www.macrotrends.net} & Daily \\ \hline
    TED Spread & \url{https://www.macrotrends.net} & Daily \\ \hline
    Trade Balance \% of GDP & \url{https://www.macrotrends.net} & Annual \\ \hline
    Debt-to-GDP Ratio & \url{https://www.macrotrends.net} & Monthly  \\ \hline 
\end{tabular}}
\end{center}
\caption{Macroeconomic Variables Considered for Finance Sector}
\label{table:finance_sector}
\end{table}

\begin{table}[htbp]
\vskip\baselineskip 
\begin{center}
\scalebox{0.8}{
\begin{tabular}{|l|c|c|} \hline
\textbf{Variables}& \textbf{Source} & \textbf{Frequency} \\ \hline
    U.S. Inflation Rate &  \url{https://www.macrotrends.net} & Annual \\ \hline
    Gold & \url{https://www.gold.org/goldhub/data/gold-prices} & Monthly \\ \hline
    Aluminum & \url{https://www.indexmundi.com/commodities} & Monthly \\ \hline
    Copper & \url{https://www.macrotrends.net} & Daily \\ \hline
    Hard Logs & \url{https://www.indexmundi.com/commodities} & Monthly \\ \hline 
    Lead & \url{https://www.indexmundi.com/commodities} & Monthly \\ \hline 
    Iron Ore & \url{https://www.indexmundi.com/commodities} & Monthly \\ \hline 
    Nickel & \url{https://www.indexmundi.com/commodities} & Monthly \\ \hline 
    Palladium & \url{https://www.macrotrends.net} & Daily \\ \hline
    Platinum & \url{https://www.macrotrends.net} & Daily \\ \hline
    Potassium Chloride & \url{https://www.indexmundi.com/commodities} & Monthly \\ \hline 
    Rock Phosphate & \url{https://www.indexmundi.com/commodities} & Monthly \\ \hline 
    Rubber & \url{https://www.indexmundi.com/commodities} & Monthly \\ \hline 
    Silver & \url{https://www.indexmundi.com/commodities} & Monthly \\ \hline 
    Tin & \url{https://www.indexmundi.com/commodities} & Monthly \\ \hline 
    Triple Superphosphate & \url{https://www.indexmundi.com/commodities} & Monthly \\ \hline 
    Urea & \url{https://www.indexmundi.com/commodities} & Monthly \\ \hline 
    Zinc & \url{https://www.indexmundi.com/commodities} & Monthly \\ \hline 
\end{tabular}}
\end{center}
\caption{Macroeconomic Variables Considered for Materials Sector}
\label{table:materials_sector}
\end{table}

\begin{table}[htbp]
\vskip\baselineskip 
\begin{center}
\scalebox{0.8}{
\begin{tabular}{|l|c|c|} \hline
\textbf{Variables}& \textbf{Source} & \textbf{Frequency} \\ \hline
    Industrial Production Index &\url{https://fred.stlouisfed.org}& Monthly \\ \hline
    Crude Oil Price &\url{https://fred.stlouisfed.org}& Monthly \\ \hline
    Capacity Utilization &\url{https://fred.stlouisfed.org}& Monthly \\ \hline
    Manufacturing &\url{https://fred.stlouisfed.org}& Monthly \\ \hline
\end{tabular}}
\end{center}
\caption{Macroeconomic Variables Considered for Industrials Sector}
\label{table:industrials_sector}
\end{table}

\begin{table}[htbp]
\vskip\baselineskip 
\begin{center}
\begin{tabular}{|l|c|c|} \hline
\textbf{Variables}& \textbf{Source} & \textbf{Frequency} \\ \hline
    Consumer Confidence Index &\url{https://data.oecd.org}& Monthly \\ \hline
    Business Confidence Index &\url{https://data.oecd.org}& Monthly \\ \hline
\end{tabular}
\end{center}
\caption{Macroeconomic Variables Considered for Consumer Goods Sector}
\label{table:consumergoods_sector}
\end{table}

\begin{table}[htbp]
\vskip\baselineskip 
\begin{center}
\scalebox{0.8}{
\begin{tabular}{|l|c|c|} \hline
\textbf{Variables}& \textbf{Source} & \textbf{Frequency} \\ \hline
    Import &\url{https://fred.stlouisfed.org}& Monthly\\ \hline
    Export Value &\url{https://data.oecd.org}&  Annual\\ \hline
    Consumer Confidence Index &\url{https://data.oecd.org}& Monthly \\ \hline
    R\&D Value &\url{https://data.oecd.org}& Annual \\ \hline
    Technology Investment &\url{https://data.oecd.org}&  Annual\\ \hline
\end{tabular}}
\end{center}
\caption{Macroeconomic Variables Considered for Technology Sector}
\label{table:technology_sector}
\end{table}

\begin{table}[htbp]
\vskip\baselineskip 
\begin{center}
\scalebox{0.8}{
\begin{tabular}{|l|c|c|} \hline
\textbf{Variables}& \textbf{Source} & \textbf{Frequency} \\ \hline
    Crude Oil Price &\url{https://fred.stlouisfed.org}& Monthly \\ \hline
    Refinery Utilization & \url{https://www.eia.gov}& Weekly \\ \hline
    Primary Energy Production &\url{https://www.eia.gov}&  Monthly\\ \hline
    Primary Energy Consumption &\url{https://www.eia.gov}&  Monthly\\ \hline
    Import &\url{https://fred.stlouisfed.org}&  Monthly\\ \hline
\end{tabular}}
\end{center}
\caption{Macroeconomic Variables Considered for Energy Sector}
\label{table:energy_sector}
\end{table}

\begin{table}[htbp]
\vskip\baselineskip 
\begin{center}
\scalebox{0.8}{
\begin{tabular}{|l|c|c|} \hline
\textbf{Variables}& \textbf{Source} & \textbf{Frequency} \\ \hline
    Crude Oil Price &\url{https://fred.stlouisfed.org}& Monthly \\ \hline
    Refinery Utilization & \url{https://www.eia.gov}& Weekly \\ \hline
    Import &\url{https://fred.stlouisfed.org}& Monthly \\ \hline
    Natural Gas Consumption &\url{https://www.eia.gov}& Annual \\ \hline
    Natural Gas Price &\url{https://www.eia.gov} & Annual \\ \hline
    Interest Rate &\url{https://fred.stlouisfed.org}&  Daily\\ \hline
    Energy Consumption &\url{https://www.eia.gov}& Monthly\\ \hline
    Electricity and Gas Production &\url{https://www.eia.gov}& Monthly\\ \hline
\end{tabular}}
\end{center}
\caption{Macroeconomic Variables Considered for Utilities Sector}
\label{table:utilities_sector}
\end{table}
\FloatBarrier

\section{Performance Plots for Different Prediction Horizons} \label{backtesting_sectors}
\subsection{Next Month Prediction Performance Plots} \label{backtesting_one_month}
\begin{figure}[!htbp]
\captionsetup{font=scriptsize,labelfont=scriptsize}
   \begin{minipage}{0.5\textwidth}
   \hspace*{-1.5cm}
     \centering
     \includegraphics[scale = 0.35]{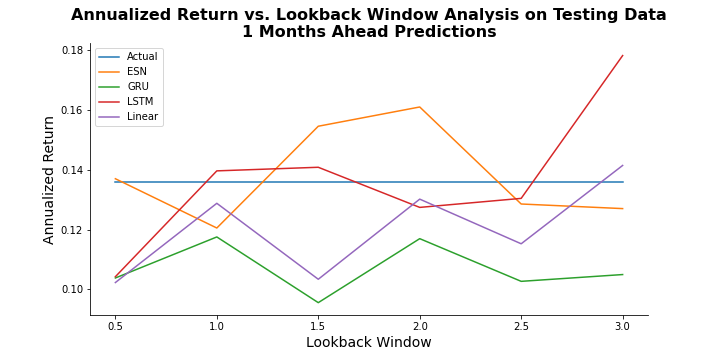}
   \end{minipage}\hfill
   \begin{minipage}{0.5\textwidth}
   \hspace*{0.5cm}
     \centering
     \includegraphics[scale = 0.35]{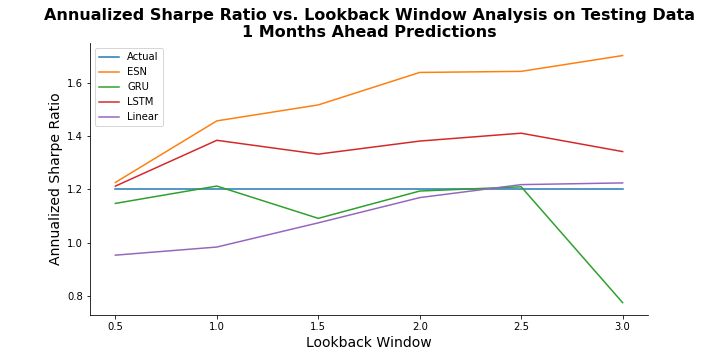}
   \end{minipage}\hfill
   \begin{minipage}{\textwidth}
     \centering
     \includegraphics[scale = 0.35]{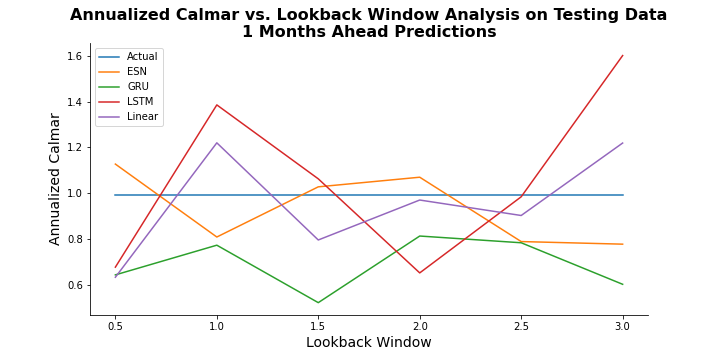}
   \end{minipage}
 \caption{Comparison of Different Lookback Windows and Different Models for Next Month Prediction (Testing Period)} \label{fig:performance_training_1_a}
\end{figure}

\begin{figure}[!htbp]
\captionsetup{font=scriptsize,labelfont=scriptsize}
   \begin{minipage}{0.5\textwidth}
   \hspace*{-1.5cm}
     \centering
     \includegraphics[scale = 0.35]{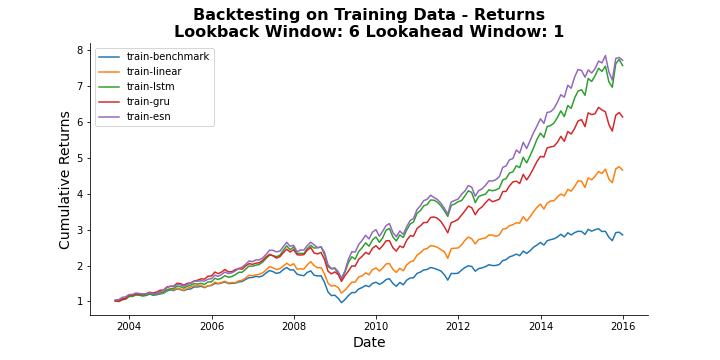}
   \end{minipage}\hfill
   \begin{minipage}{0.5\textwidth}
   \hspace*{0.5cm}
     \centering
     \includegraphics[scale = 0.35]{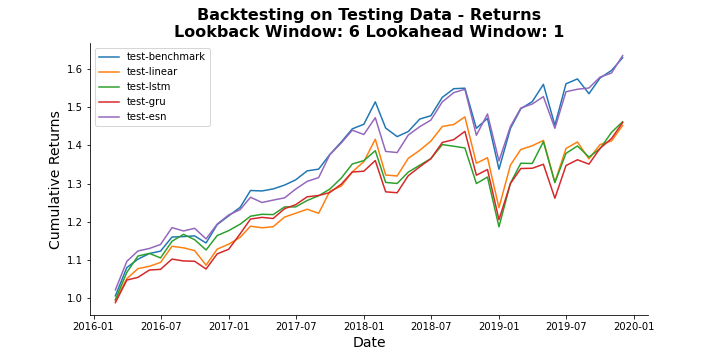}
   \end{minipage}
 \caption{Backtesting Performance for Next Month Prediction Models (6 Months Lookback Window)} \label{fig:backtesting_1_6}
\end{figure}

\begin{figure}[!htbp]
\captionsetup{font=scriptsize,labelfont=scriptsize}
   \begin{minipage}{0.5\textwidth}
   \hspace*{-1.5cm}
     \centering
     \includegraphics[scale = 0.35]{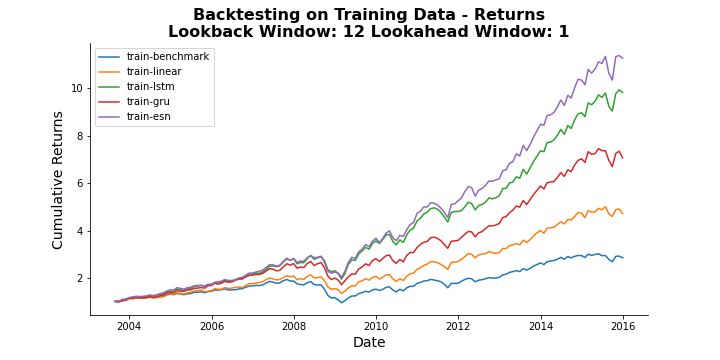}
   \end{minipage}\hfill
   \begin{minipage}{0.5\textwidth}
   \hspace*{0.5cm}
     \centering
     \includegraphics[scale = 0.35]{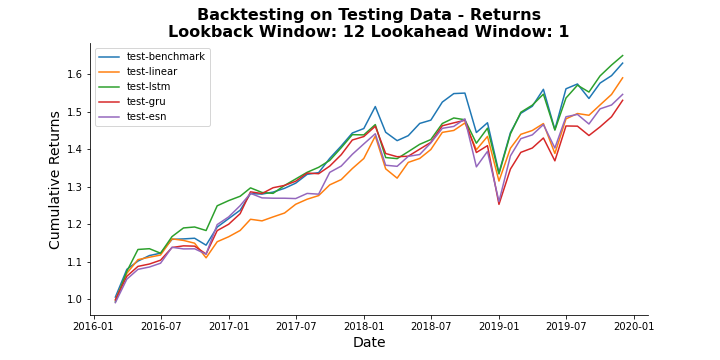}
   \end{minipage}
 \caption{Backtesting Performance for Next Month Prediction Models (One Year Lookback Window)} \label{fig:backtesting_1_12}
\end{figure}

\begin{figure}[!htbp]
\captionsetup{font=scriptsize,labelfont=scriptsize}
   \begin{minipage}{0.5\textwidth}
   \hspace*{-1.5cm}
     \centering
     \includegraphics[scale = 0.35]{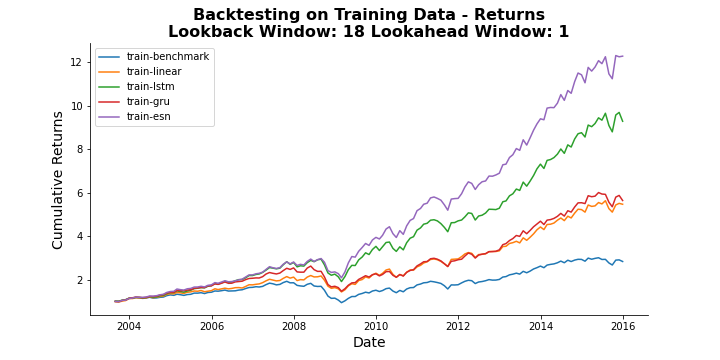}
   \end{minipage}\hfill
   \begin{minipage}{0.5\textwidth}
   \hspace*{0.5cm}
     \centering
     \includegraphics[scale = 0.35]{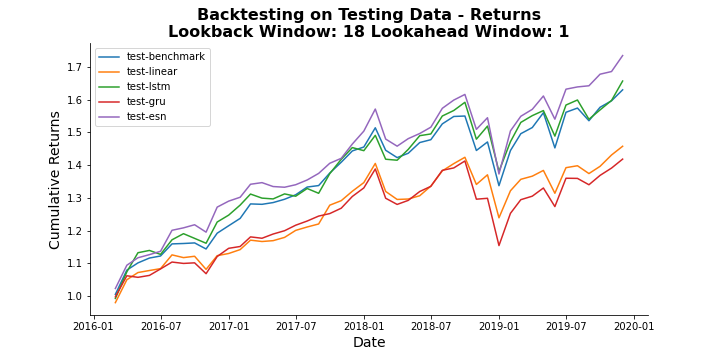}
   \end{minipage}
 \caption{Backtesting Performance for Next Month Prediction Models (1.5 Years Lookback Window)} \label{fig:backtesting_1_18}
\end{figure}

\begin{figure}[!htbp]
\captionsetup{font=scriptsize,labelfont=scriptsize}
   \begin{minipage}{0.5\textwidth}
   \hspace*{-1.5cm}
     \centering
     \includegraphics[scale = 0.35]{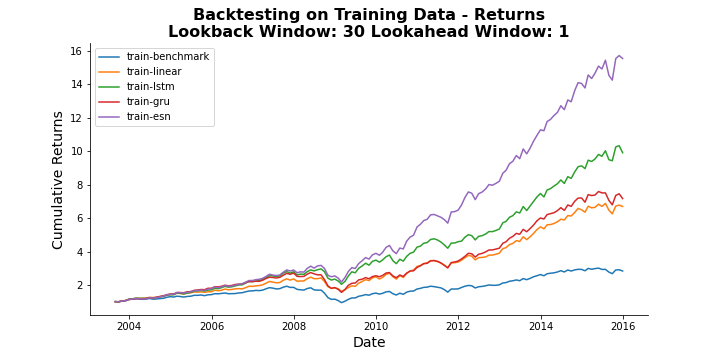}
   \end{minipage}\hfill
   \begin{minipage}{0.5\textwidth}
   \hspace*{0.5cm}
     \centering
     \includegraphics[scale = 0.35]{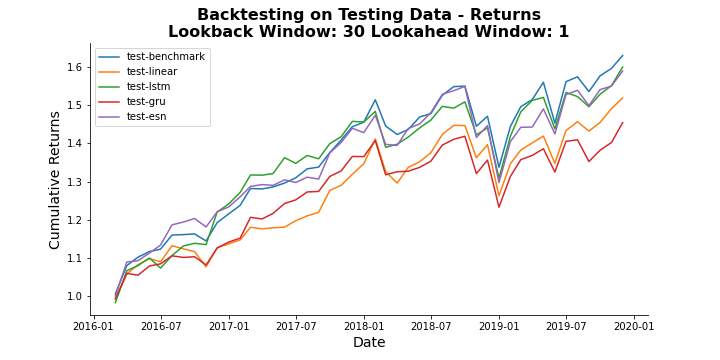}
   \end{minipage}
 \caption{Backtesting Performance for Next Month Prediction Models (2.5 Years Lookback Window)} \label{fig:backtesting_1_30}
\end{figure}

\begin{figure}[!htbp]
\captionsetup{font=scriptsize,labelfont=scriptsize}
   \begin{minipage}{0.5\textwidth}
   \hspace*{-1.5cm}
     \centering
     \includegraphics[scale = 0.35]{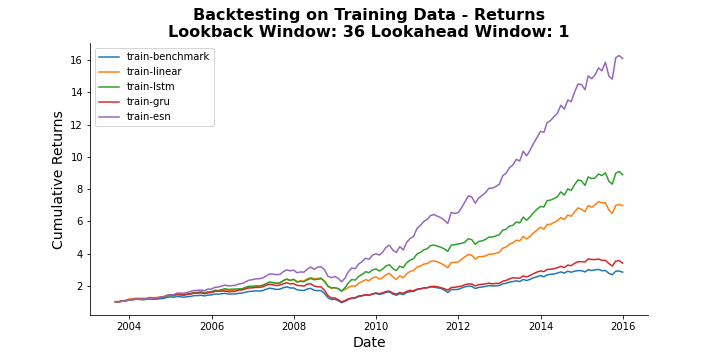}
   \end{minipage}\hfill
   \begin{minipage}{0.5\textwidth}
   \hspace*{0.5cm}
     \centering
     \includegraphics[scale = 0.35]{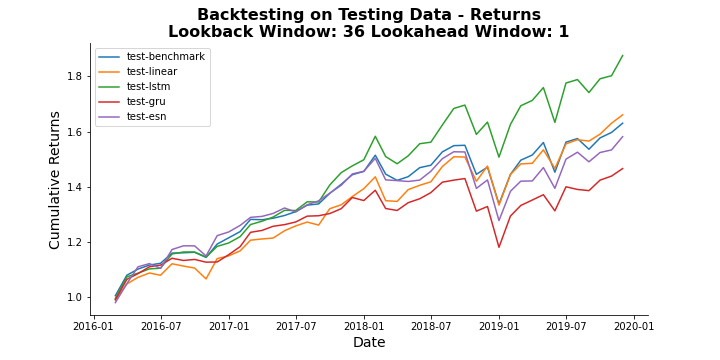}
   \end{minipage}
 \caption{Backtesting Performance for Next Month Prediction Models (Three Years Lookback Window)} \label{fig:backtesting_1_36}
\end{figure}

Figure \ref{fig:performance_training_1_a} illustrates how different models perform across different lookback windows during testing period. According to annualized Sharpe ratios, \texttt{ESN}s perform better than other models, and \texttt{LSTM} follows \texttt{ESN}. During the testing period, the annualized Sharpe ratio increases as more historical data is used for \texttt{ESN} model. There is no general pattern for annualized return and annualized Calmar ratio, but the best performing model for each lookback window is either \texttt{LSTM} or \texttt{ESN}.
\FloatBarrier
Figures \ref{fig:backtesting_1_6}, \ref{fig:backtesting_1_12}, \ref{fig:backtesting_1_18}, \ref{fig:backtesting_1_30}, and \ref{fig:backtesting_1_36} illustrate the backtesting results for one month ahead ranking prediction with six months, one year, 1.5 years, 2.5 years, and three years, respectively. According to the plots, \texttt{ESN} constantly perform the best during the training period, and all four models are generating better returns than the benchmark portfolio. During the testing period, the results are not as obvious as the results obtained for training period. However, it can be observed that \texttt{ESN} and \texttt{LSTM} models performed better than \texttt{GRU} and Ridge regression models. 
\FloatBarrier

\FloatBarrier

\FloatBarrier

\FloatBarrier

\FloatBarrier

\subsection{Three Months Ahead Prediction Performance Plots} \label{backtesting_three_months}
\begin{figure}[!htbp]
\captionsetup{font=scriptsize,labelfont=scriptsize}
   \begin{minipage}{0.5\textwidth}
   \hspace*{-1.5cm}
     \centering
     \includegraphics[scale = 0.35]{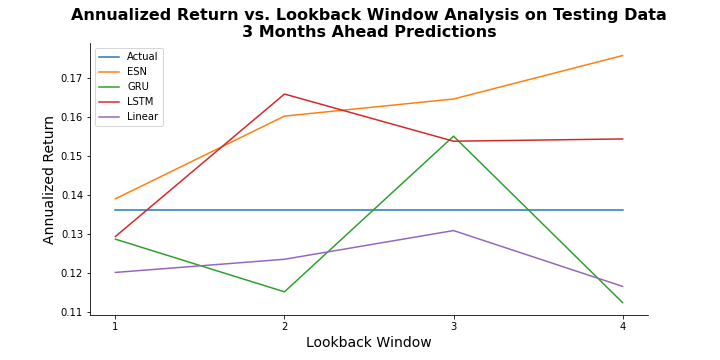}
   \end{minipage}\hfill
   \begin{minipage}{0.5\textwidth}
   \hspace*{0.5cm}
     \centering
     \includegraphics[scale = 0.35]{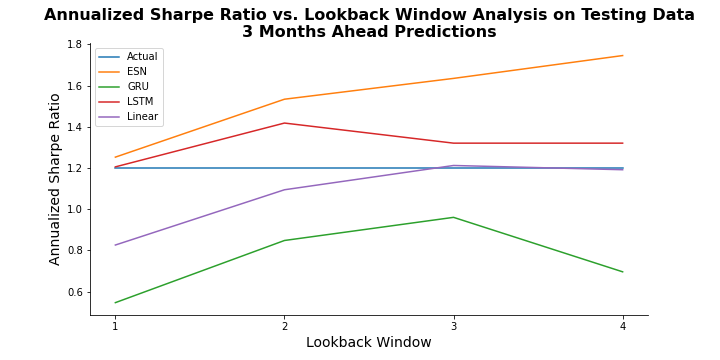}
   \end{minipage}\hfill
   \begin{minipage}{\textwidth}
     \centering
     \includegraphics[scale = 0.35]{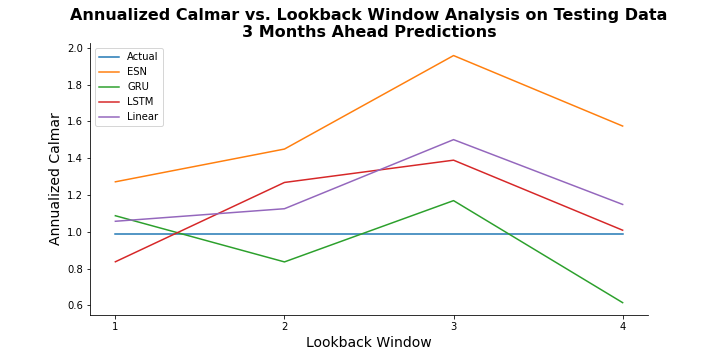}
   \end{minipage}
 \caption{Comparison of Different Lookback Windows and Different Models for Three Months Ahead Prediction (Testing Period)} \label{fig:performance_training_3_a}
\end{figure}

\begin{figure}[!htbp]
\captionsetup{font=scriptsize,labelfont=scriptsize}
   \begin{minipage}{0.5\textwidth}
   \hspace*{-1.5cm}
     \centering
     \includegraphics[scale = 0.35]{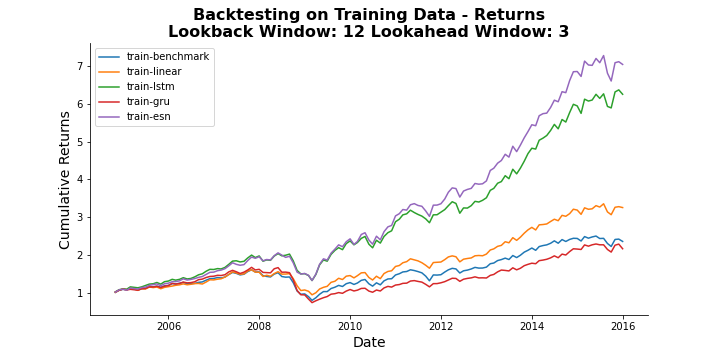}
   \end{minipage}\hfill
   \begin{minipage}{0.5\textwidth}
   \hspace*{0.5cm}
     \centering
     \includegraphics[scale = 0.35]{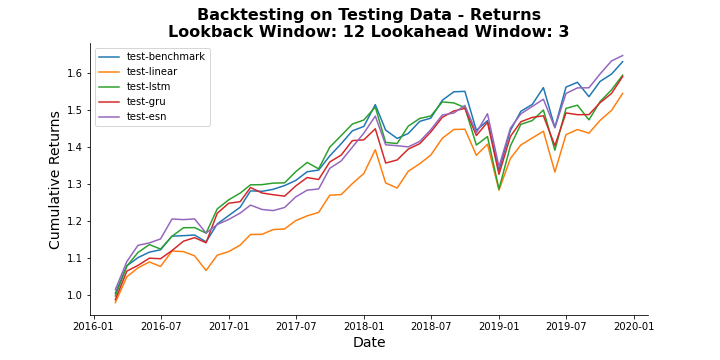}
   \end{minipage}
 \caption{Backtesting Performance for Three Months Ahead Prediction Models (One Year Lookback Window)} \label{fig:backtesting_3_12}
\end{figure}

\begin{figure}[!htbp]
\captionsetup{font=scriptsize,labelfont=scriptsize}
   \begin{minipage}{0.5\textwidth}
   \hspace*{-1.5cm}
     \centering
     \includegraphics[scale = 0.35]{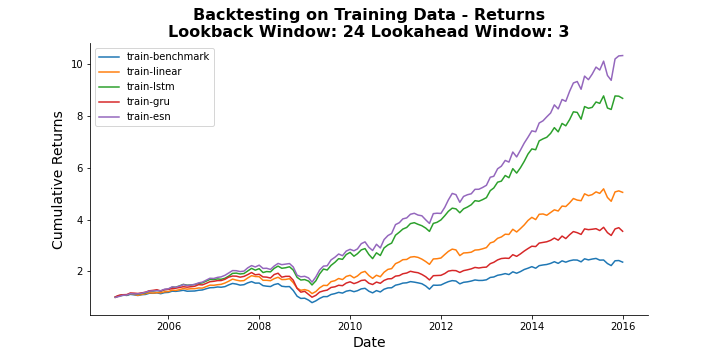}
   \end{minipage}\hfill
   \begin{minipage}{0.5\textwidth}
   \hspace*{0.5cm}
     \centering
     \includegraphics[scale = 0.35]{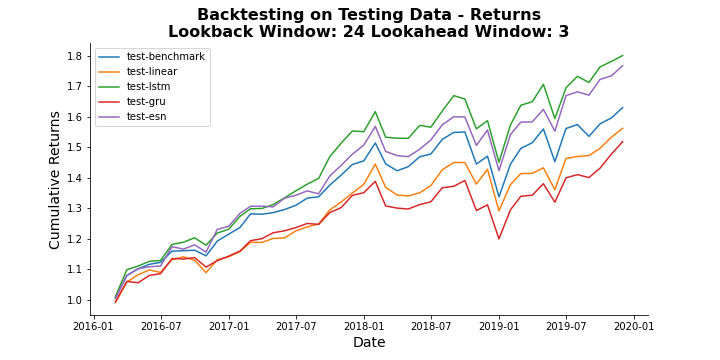}
   \end{minipage}
 \caption{Backtesting Performance for Three Months Ahead Prediction Models (Two Years Lookback Window)} \label{fig:backtesting_3_24}
\end{figure}

\begin{figure}[!htbp]
\captionsetup{font=scriptsize,labelfont=scriptsize}
   \begin{minipage}{0.5\textwidth}
   \hspace*{-1.5cm}
     \centering
     \includegraphics[scale = 0.35]{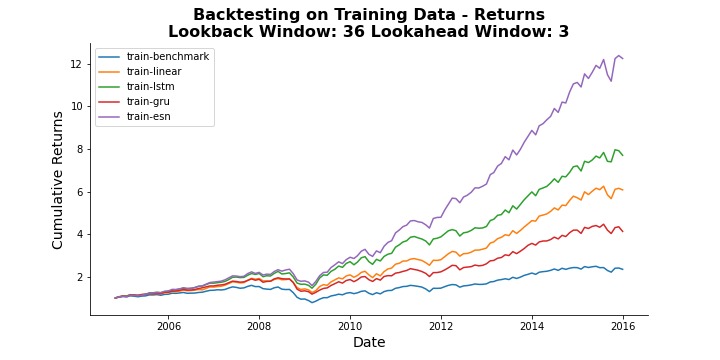}
   \end{minipage}\hfill
   \begin{minipage}{0.5\textwidth}
   \hspace*{0.5cm}
     \centering
     \includegraphics[scale = 0.35]{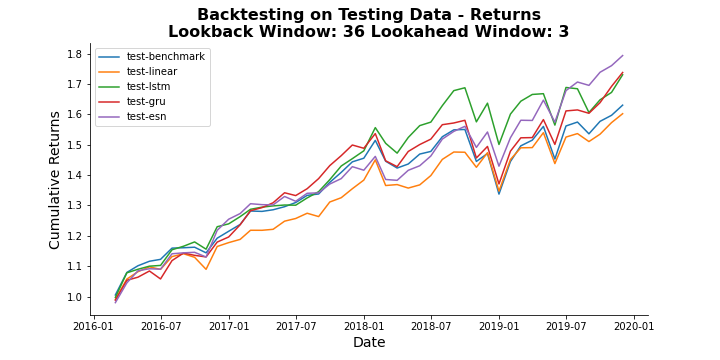}
   \end{minipage}
 \caption{Backtesting Performance for Three Months Ahead Prediction Models (Three Years Lookback Window)} \label{fig:backtesting_3_36}
\end{figure}

Figure \ref{fig:performance_training_3_a} illustrates how different models behave using different lookback windows during testing period for three months ahead predictions. There is no generic behavior obtained from all models, but \texttt{ESN} performs the best in terms of all performance metrics compared to other three models and also benchmark portfolio. Annualized return and annualized Sharpe ratio obtained from \texttt{ESN} models increase as the lookback window increases. For annualized Calmar ratio, the performance of all models increases until three years of historical data, but addition of one more year of data results in a decrease in the performance. 
\FloatBarrier
Figures \ref{fig:backtesting_3_12}, \ref{fig:backtesting_3_24}, and \ref{fig:backtesting_3_36} provides the backtesting result for three months ahead predictions using one year, two years and three years of historical data, respectively. The results show that \texttt{ESN} outperforms all other models during training period, and \texttt{LSTM} follows \texttt{ESN} closely for one year and two years of historical data. Testing period plots show that \texttt{ESN} and \texttt{LSTM} models obtain higher cumulative returns than other models and benchmark portfolio most of the time. However, the difference between performance of different models is not as clear as the difference obtained during training horizon. 
\FloatBarrier

\FloatBarrier

\FloatBarrier

\subsection{Six Months Ahead Prediction Performance Plots}
\begin{figure}[!htbp]
\captionsetup{font=scriptsize,labelfont=scriptsize}
   \begin{minipage}{0.5\textwidth}
   \hspace*{-1.5cm}
     \centering
     \includegraphics[scale = 0.35]{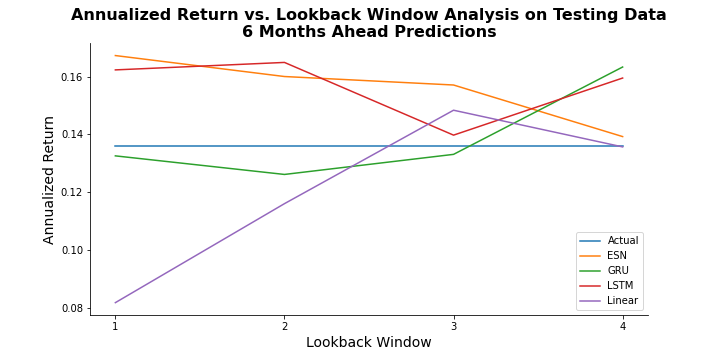}
   \end{minipage}\hfill
   \begin{minipage}{0.5\textwidth}
   \hspace*{0.5cm}
     \centering
     \includegraphics[scale = 0.35]{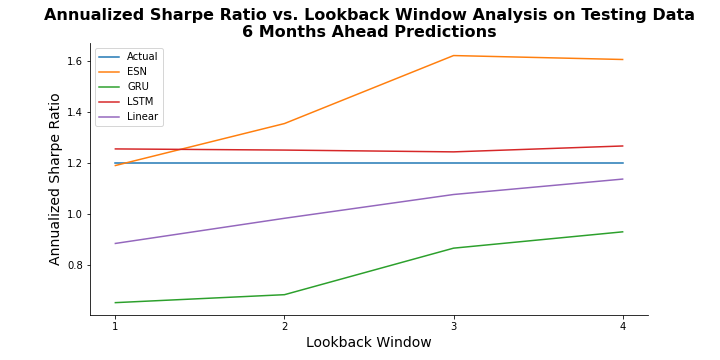}
   \end{minipage}\hfill
   \begin{minipage}{\textwidth}
     \centering
     \includegraphics[scale = 0.35]{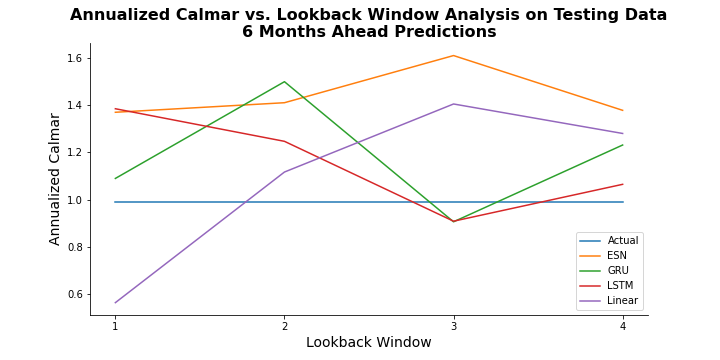}
   \end{minipage}
 \caption{Comparison of Different Lookback Windows and Different Models for Six Months Ahead Prediction (Testing Period)} \label{fig:performance_training_6_a}
\end{figure}

Figure \ref{fig:performance_training_6_a} illustrates the performance metrics plots for six months ahead prediction models during testing period. The plots show that \texttt{ESN} model generally performs the best for each performance metric and lookback window. Although annualized Sharpe ratio decreases after three years of historical data for \texttt{ESN} model, Sharpe ratio tends to increase as the lookback window increases for all models. For annualized returns and annualized Calmar ratio, \texttt{ESN} and ridge regression follow similar patterns. The same behavior is observed between \texttt{LSTM} and \texttt{GRU} models.

\begin{figure}[!htbp]
\captionsetup{font=scriptsize,labelfont=scriptsize}
   \begin{minipage}{0.5\textwidth}
   \hspace*{-1.5cm}
     \centering
     \includegraphics[scale = 0.35]{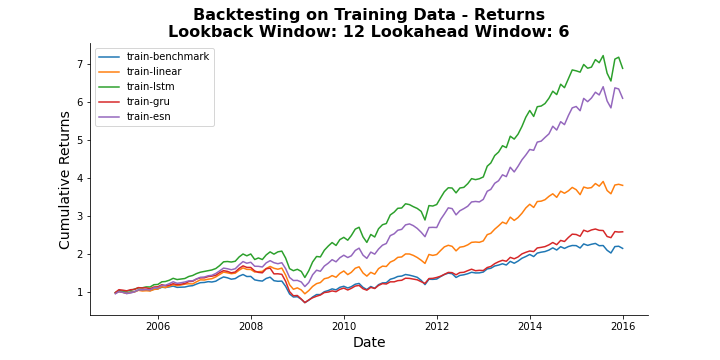}
   \end{minipage}\hfill
   \begin{minipage}{0.5\textwidth}
   \hspace*{0.5cm}
     \centering
     \includegraphics[scale = 0.35]{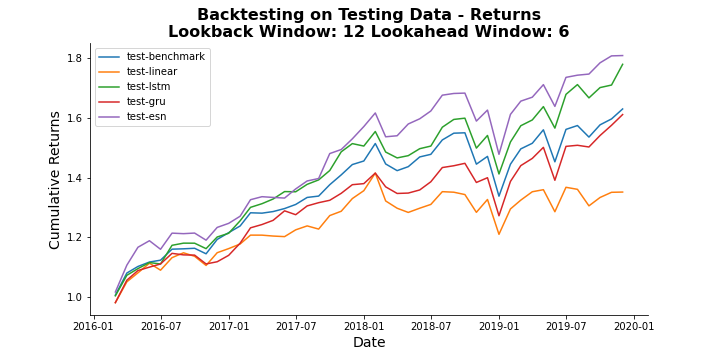}
   \end{minipage}
 \caption{Backtesting Performance for Six Months Ahead Prediction Models (One Year Lookback Window)} \label{fig:backtesting_6_12}
\end{figure}

\begin{figure}[!htbp]
\captionsetup{font=scriptsize,labelfont=scriptsize}
   \begin{minipage}{0.5\textwidth}
   \hspace*{-1.5cm}
     \centering
     \includegraphics[scale = 0.35]{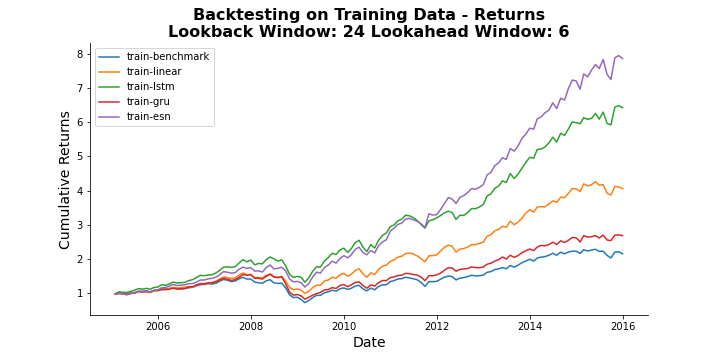}
   \end{minipage}\hfill
   \begin{minipage}{0.5\textwidth}
   \hspace*{0.5cm}
     \centering
     \includegraphics[scale = 0.35]{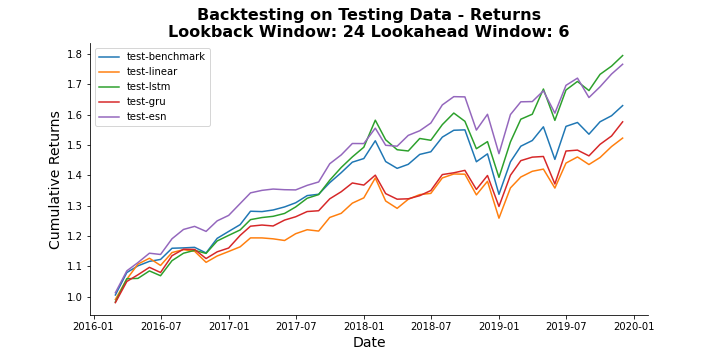}
   \end{minipage}
 \caption{Backtesting Performance for Six Months Ahead Prediction Models (Two Years Lookback Window)} \label{fig:backtesting_6_24}
\end{figure}

\begin{figure}[!htbp]
\captionsetup{font=scriptsize,labelfont=scriptsize}
   \begin{minipage}{0.5\textwidth}
   \hspace*{-1.5cm}
     \centering
     \includegraphics[scale = 0.35]{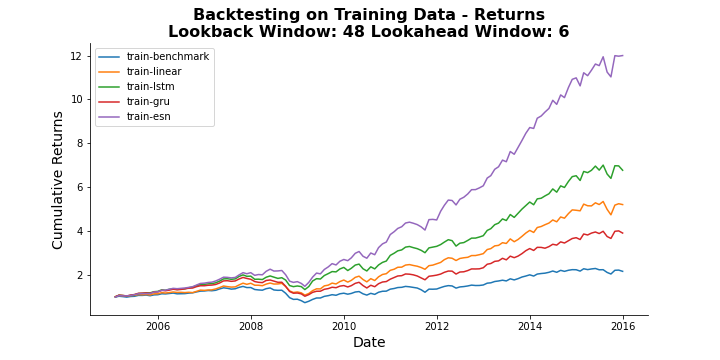}
   \end{minipage}\hfill
   \begin{minipage}{0.5\textwidth}
   \hspace*{0.5cm}
     \centering
     \includegraphics[scale = 0.35]{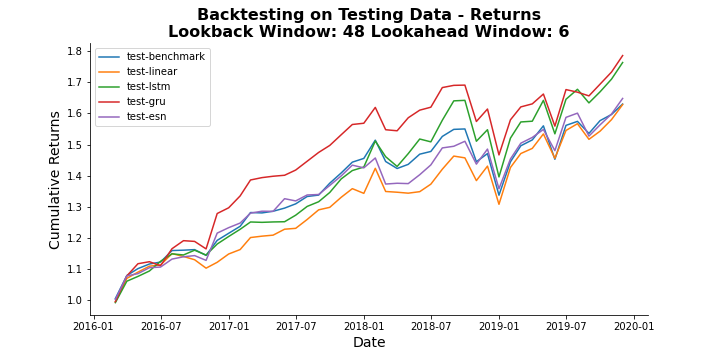}
   \end{minipage}
 \caption{Backtesting Performance for Six Months Ahead Prediction Models (Four Years Lookback Window)} \label{fig:backtesting_6_48}
\end{figure}

\label{backtesting_six_months}
\FloatBarrier
Figures \ref{fig:backtesting_6_12}, \ref{fig:backtesting_6_24}, and \ref{fig:backtesting_6_48} illustrate the backtesting results for six months ahead prediction models. The results are consistent with near-term backtesting results. All models perform better than benchmark portfolio over training horizon, and \texttt{ESN} performs the best for each lookback window. Although \texttt{ESN} performs the best for one year and two years lookback windows during testing period, \texttt{GRU} outperforms all other models when four years of historical data is used. \texttt{LSTM} model perfoms the second best for all lookback windows within testing period.  
\FloatBarrier

\FloatBarrier

\FloatBarrier

\subsection{One Year Ahead Prediction Performance Plots} \label{backtesting_one_year}

\begin{figure}[!htbp]
\captionsetup{font=scriptsize,labelfont=scriptsize}
   \begin{minipage}{0.5\textwidth}
   \hspace*{-1.5cm}
     \centering
     \includegraphics[scale = 0.35]{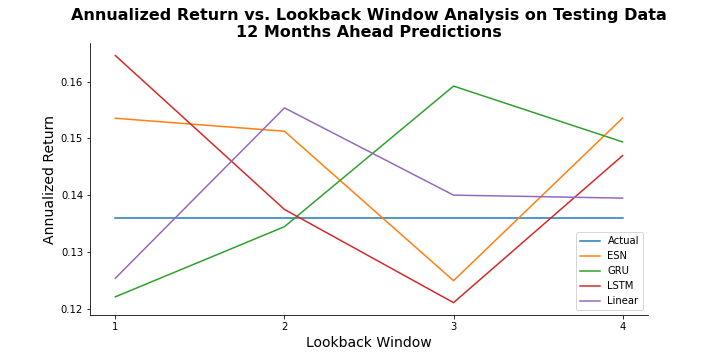}
   \end{minipage}\hfill
   \begin{minipage}{0.5\textwidth}
   \hspace*{0.5cm}
     \centering
     \includegraphics[scale = 0.35]{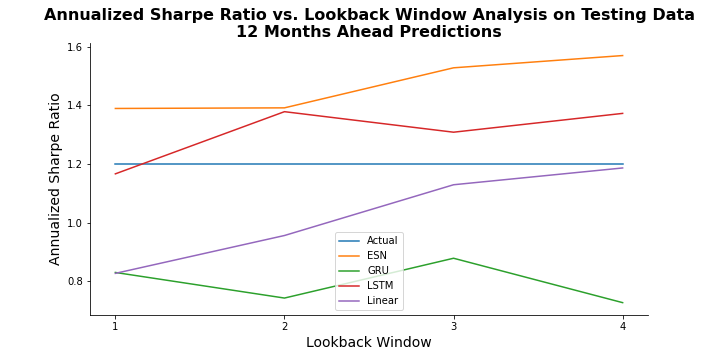}
   \end{minipage}\hfill
   \begin{minipage}{\textwidth}
     \centering
     \includegraphics[scale = 0.35]{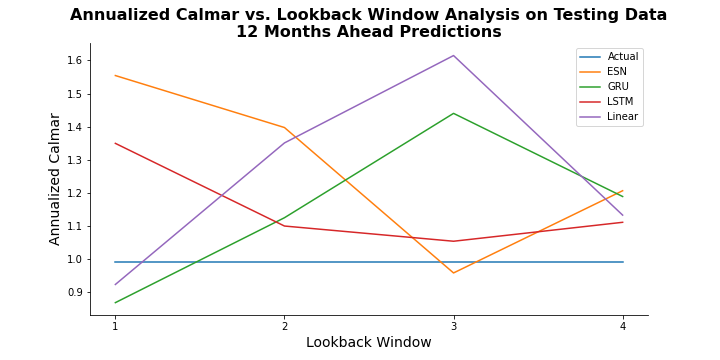}
   \end{minipage}
 \caption{Comparison of Different Lookback Windows and Different Models for One Year Ahead Prediction (Testing Period)} \label{fig:performance_training_12_a}
\end{figure}

Figure \ref{fig:performance_training_12_a} illustrates the performance of different models with different lookback windows over testing period for one year ahead sector ranking predictions. Both annualized return and annualized Calmar ratio plots reveal that the performance of \texttt{ESN} and \texttt{LSTM} models decrease until three years of historical data, and then it starts increasing. \texttt{GRU} and ridge regression models follow the similar patterns yet their behavior is opposite of \texttt{ESN} and \texttt{LSTM}. Annualized Sharpe ratio results indicate that \texttt{ESN} is superior to other models during testing period. Sharpe ratio values tend to increase for \texttt{ESN} and ridge regression models as lookback window increases. 

\begin{figure}[!htbp]
\captionsetup{font=scriptsize,labelfont=scriptsize}
   \begin{minipage}{0.5\textwidth}
   \hspace*{-1.5cm}
     \centering
     \includegraphics[scale = 0.35]{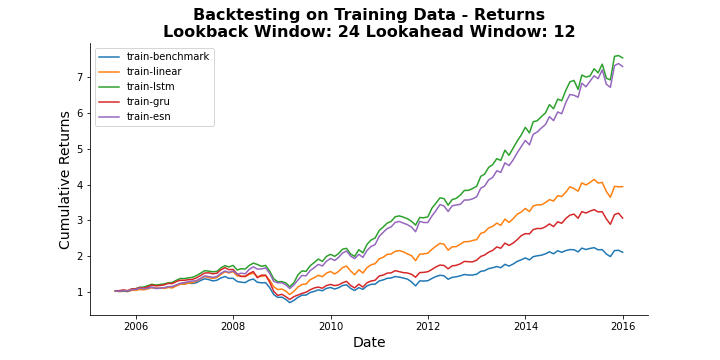}
   \end{minipage}\hfill
   \begin{minipage}{0.5\textwidth}
   \hspace*{0.5cm}
     \centering
     \includegraphics[scale = 0.35]{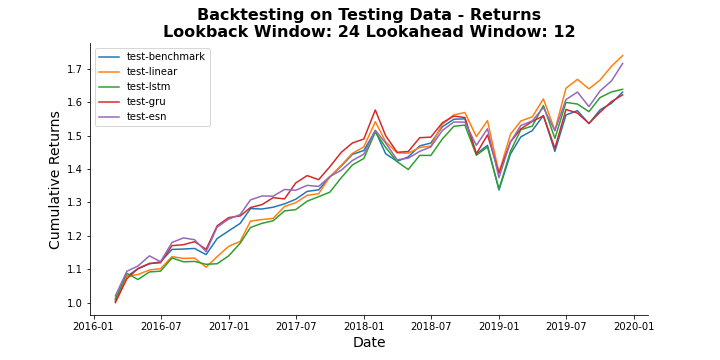}
   \end{minipage}
 \caption{Backtesting Performance for One Year Ahead Prediction Models (Two Years Lookback Window)} \label{fig:backtesting_12_24}
\end{figure}

\begin{figure}[!htbp]
\captionsetup{font=scriptsize,labelfont=scriptsize}
   \begin{minipage}{0.5\textwidth}
   \hspace*{-1.5cm}
     \centering
     \includegraphics[scale = 0.35]{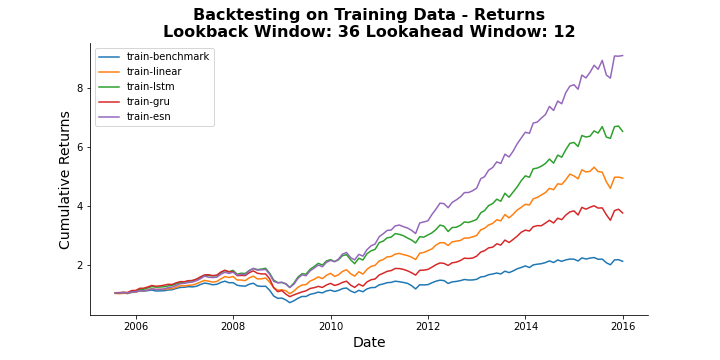}
   \end{minipage}\hfill
   \begin{minipage}{0.5\textwidth}
   \hspace*{0.5cm}
     \centering
     \includegraphics[scale = 0.35]{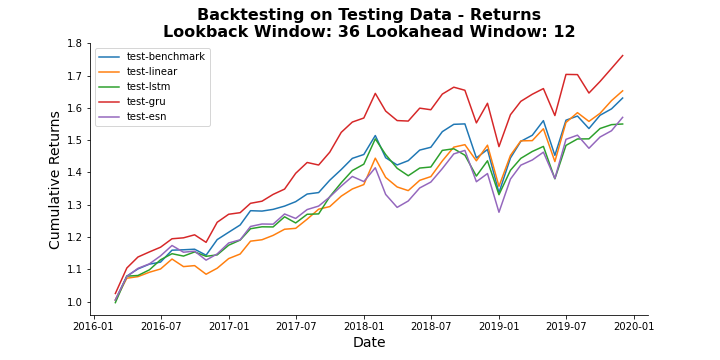}
   \end{minipage}
 \caption{Backtesting Performance for One Year Ahead Prediction Models (Three Years Lookback Window)} \label{fig:backtesting_12_36}
\end{figure}

\begin{figure}[!htbp]
\captionsetup{font=scriptsize,labelfont=scriptsize}
   \begin{minipage}{0.5\textwidth}
   \hspace*{-1.5cm}
     \centering
     \includegraphics[scale = 0.35]{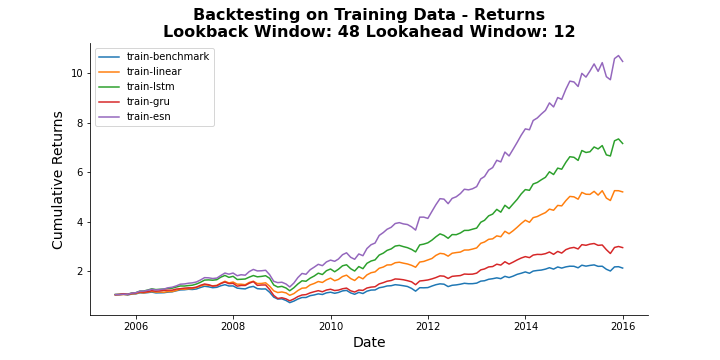}
   \end{minipage}\hfill
   \begin{minipage}{0.5\textwidth}
   \hspace*{0.5cm}
     \centering
     \includegraphics[scale = 0.35]{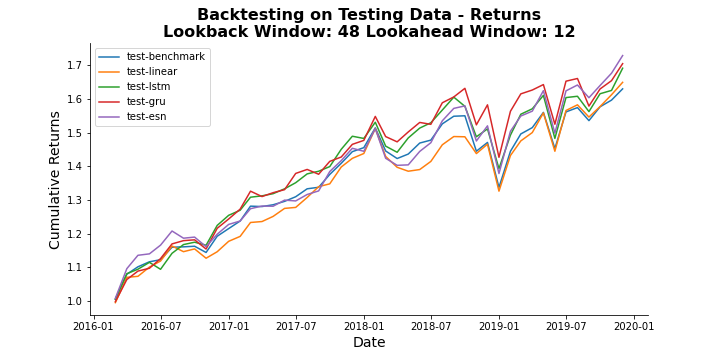}
   \end{minipage}
 \caption{Backtesting Performance for One Year Ahead Prediction Models (Four Years Lookback Window)} \label{fig:backtesting_12_48}
\end{figure}

\FloatBarrier
Figures \ref{fig:backtesting_12_24}, \ref{fig:backtesting_12_36}, and \ref{fig:backtesting_12_48} provides the backtesting results for one month ahead predictions with two years, three years, and four years of historical data, respectively. During the training period, all models perform better than benchmark portfolio, and \texttt{ESN} achieves highest cumulative returns when two years or three years of historical data is used. Although \texttt{ESN} generally performs better than benchmark portfolio, the results show that \texttt{GRU} model outperforms all other models when three or four years of historical data is used. 
\FloatBarrier

\FloatBarrier

\FloatBarrier

\subsection{Two Years Ahead Prediction Performance Plots}
\label{backtesting_two_years}

\begin{figure}[!htbp]
\captionsetup{font=scriptsize,labelfont=scriptsize}
   \begin{minipage}{0.5\textwidth}
   \hspace*{-1.5cm}
     \centering
     \includegraphics[scale = 0.35]{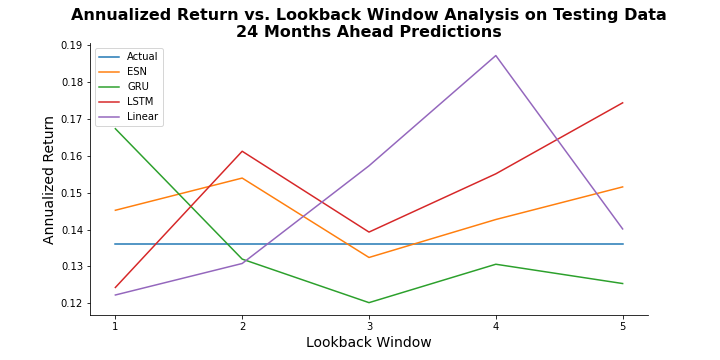}
   \end{minipage}\hfill
   \begin{minipage}{0.5\textwidth}
   \hspace*{0.5cm}
     \centering
     \includegraphics[scale = 0.35]{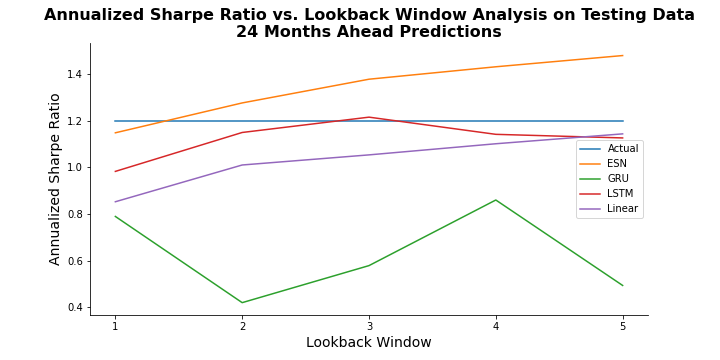}
   \end{minipage}\hfill
   \begin{minipage}{\textwidth}
     \centering
     \includegraphics[scale = 0.35]{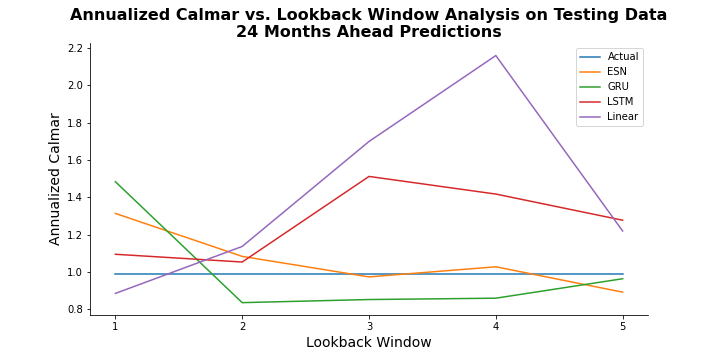}
   \end{minipage}
 \caption{Comparison of Different Lookback Windows and Different Models for Two Years Ahead Prediction (Testing Period)} \label{fig:performance_training_24_a}
\end{figure}

Figure \ref{fig:performance_training_24_a} gives an overview of the performance of two years ahead prediction models. \texttt{ESN} outperforms other models in terms annualized Sharpe ratio, and the annualized Sharpe ratio has a positive relationship with length of the lookback window for \texttt{ESN} and ridge regression models. Annualized return and annualized Calmar ratio plots show that \texttt{GRU} performs the best when the lookback window is one year, but then ridge regression and \texttt{LSTM} obtains higher values in comparison to \texttt{ESN} and \texttt{GRU}.

\begin{figure}[!htbp]
\captionsetup{font=scriptsize,labelfont=scriptsize}
   \begin{minipage}{0.5\textwidth}
   \hspace*{-1.5cm}
     \centering
     \includegraphics[scale = 0.35]{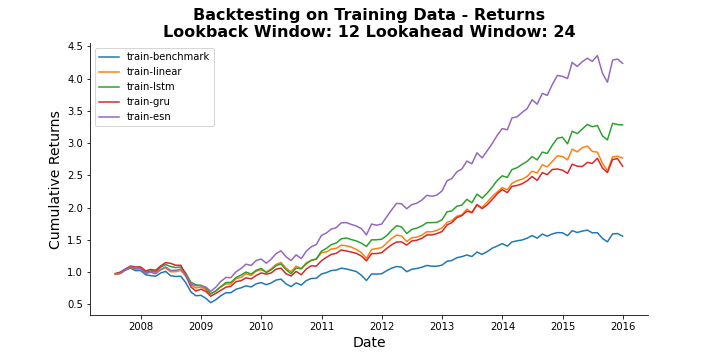}
   \end{minipage}\hfill
   \begin{minipage}{0.5\textwidth}
   \hspace*{0.5cm}
     \centering
     \includegraphics[scale = 0.35]{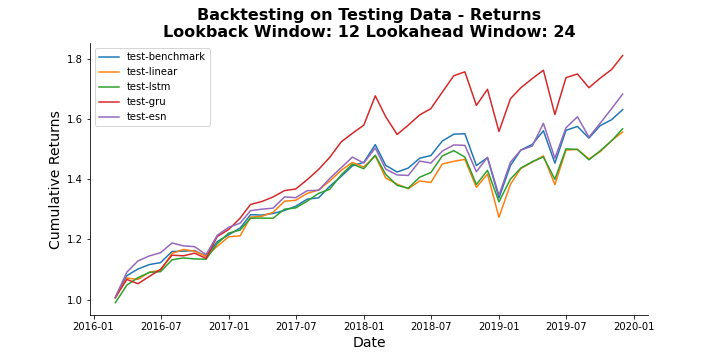}
   \end{minipage}
 \caption{Backtesting Performance for Two Years Ahead Prediction Models (One Year Lookback Window)} \label{fig:backtesting_24_12}
\end{figure}

\begin{figure}[!htbp]
\captionsetup{font=scriptsize,labelfont=scriptsize}
   \begin{minipage}{0.5\textwidth}
   \hspace*{-1.5cm}
     \centering
     \includegraphics[scale = 0.35]{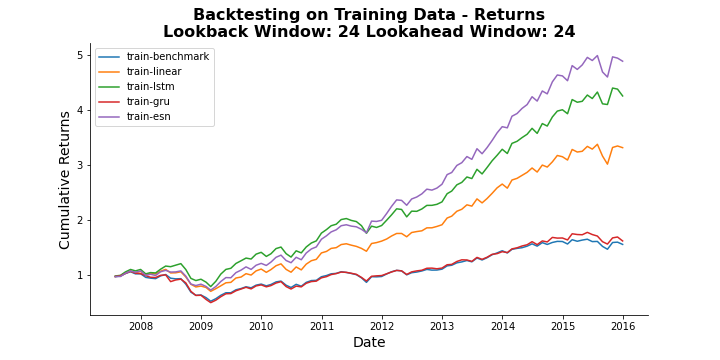}
   \end{minipage}\hfill
   \begin{minipage}{0.5\textwidth}
   \hspace*{0.5cm}
     \centering
     \includegraphics[scale = 0.35]{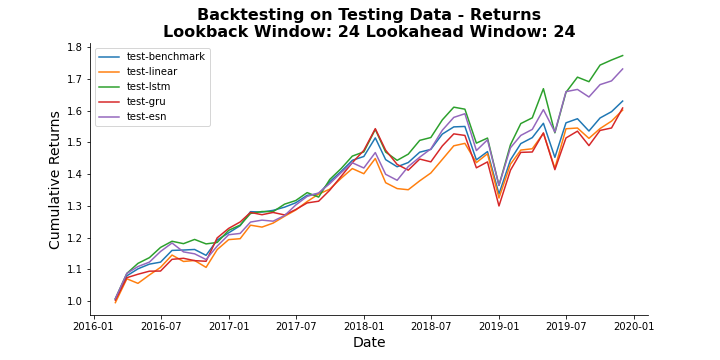}
   \end{minipage}
 \caption{Backtesting Performance for Two Years Ahead Prediction Models (Two Years Years Lookback Window)} \label{fig:backtesting_24_24}
\end{figure}

\begin{figure}[!htbp]
\captionsetup{font=scriptsize,labelfont=scriptsize}
   \begin{minipage}{0.5\textwidth}
   \hspace*{-1.5cm}
     \centering
     \includegraphics[scale = 0.35]{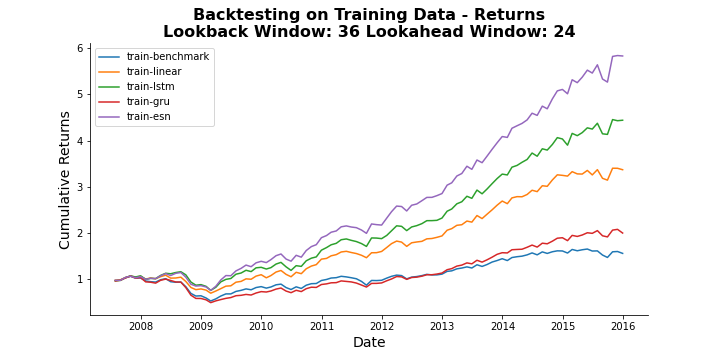}
   \end{minipage}\hfill
   \begin{minipage}{0.5\textwidth}
   \hspace*{0.5cm}
     \centering
     \includegraphics[scale = 0.35]{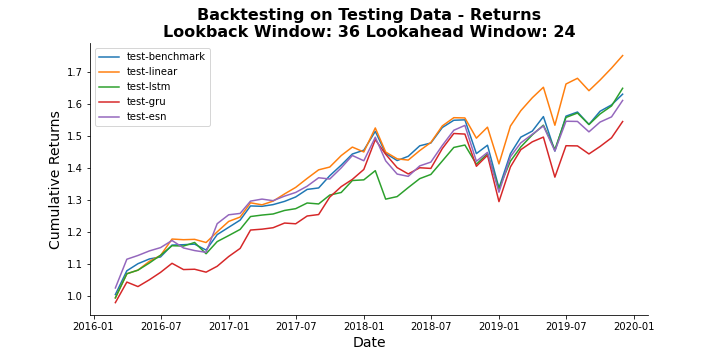}
   \end{minipage}
 \caption{Backtesting Performance for Two Years Ahead Prediction Models (Third Years Lookback Window)} \label{fig:backtesting_24_36}
\end{figure}

\begin{figure}[!htbp]
\captionsetup{font=scriptsize,labelfont=scriptsize}
   \begin{minipage}{0.5\textwidth}
   \hspace*{-1.5cm}
     \centering
     \includegraphics[scale = 0.35]{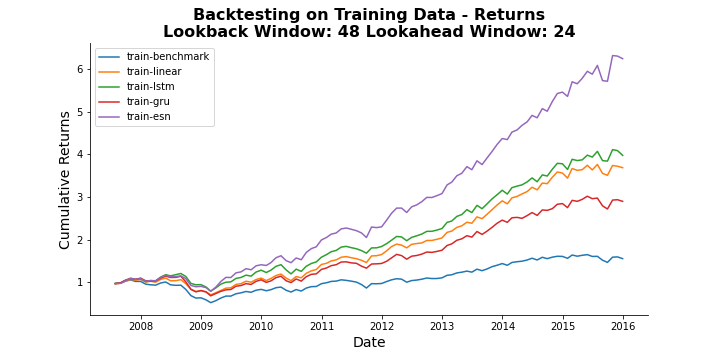}
   \end{minipage}\hfill
   \begin{minipage}{0.5\textwidth}
   \hspace*{0.5cm}
     \centering
     \includegraphics[scale = 0.35]{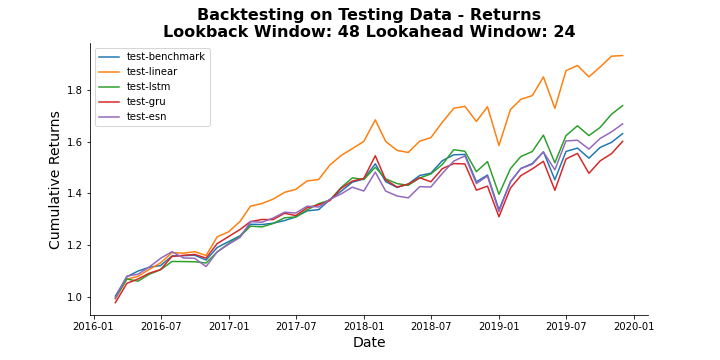}
   \end{minipage}
 \caption{Backtesting Performance for Two Years Ahead Prediction Models (Four Years Lookback Window)} \label{fig:backtesting_24_48}
\end{figure}

\FloatBarrier
Figures \ref{fig:backtesting_24_12}, \ref{fig:backtesting_24_24}, \ref{fig:backtesting_24_36}, and \ref{fig:backtesting_24_48} illustrate that \texttt{ESN} performs the best within training period. \texttt{ESN} still performs better than benchmark portfolio during testing period, but it is beaten by other models in the existence of different lookback windows.  
\FloatBarrier

\FloatBarrier

\FloatBarrier

\FloatBarrier



\end{document}